\documentclass[11pt, a4paper,onecolumn, aps]{revtex4-2}

\usepackage[margin=0.5in]{geometry}
\usepackage{amsmath, amssymb, mathtools, bm}
\usepackage{graphicx}
\usepackage{setspace}
\usepackage{float}
\usepackage{textgreek}
\usepackage{natbib}
\usepackage{caption}
\usepackage{subcaption}
\usepackage{physics}
\usepackage{siunitx}
\usepackage{multirow}
\AtBeginDocument{\RenewCommandCopy\qty\SI}
\usepackage{xcolor}
\usepackage{booktabs}
\usepackage{microtype}
\usepackage[colorlinks=true,linkcolor=blue,citecolor=red]{hyperref}
\usepackage[separate-uncertainty=true]{siunitx}
\usepackage{tikz}
\usepackage{pgfplots}
\usepackage{url}
\usepackage{bm}
\usetikzlibrary{decorations.pathmorphing}
\pgfplotsset{compat=1.18}
\usepackage[T1]{fontenc}

\usepackage[en-US]{datetime2}
\DTMlangsetup*{showdayofmonth=false} % Hide the day of the month
\DeclareUnicodeCharacter{0301}{\'{e}}
\DeclareUnicodeCharacter{211A}{$\mathbb{Q}$}
\definecolor{BLUE}{RGB}{0,0,255} % Defining BLUE as a color using RGB values

\begin{document}
\raggedbottom
\title{\Large Reconstructions of \texorpdfstring{$f(\mathcal{P})$}{Lg} and \texorpdfstring{$f(\mathcal{Q})$}{Lg} gravity models from \texorpdfstring{$(m,n)$}{Lg}-type Barrow Holographic Dark Energy: Analysis and Observational Constraints}

\author{Tamal Mukhopadhyay~\href{https://orcid.org/0000-0001-9843-906X}{\includegraphics[height=1em]{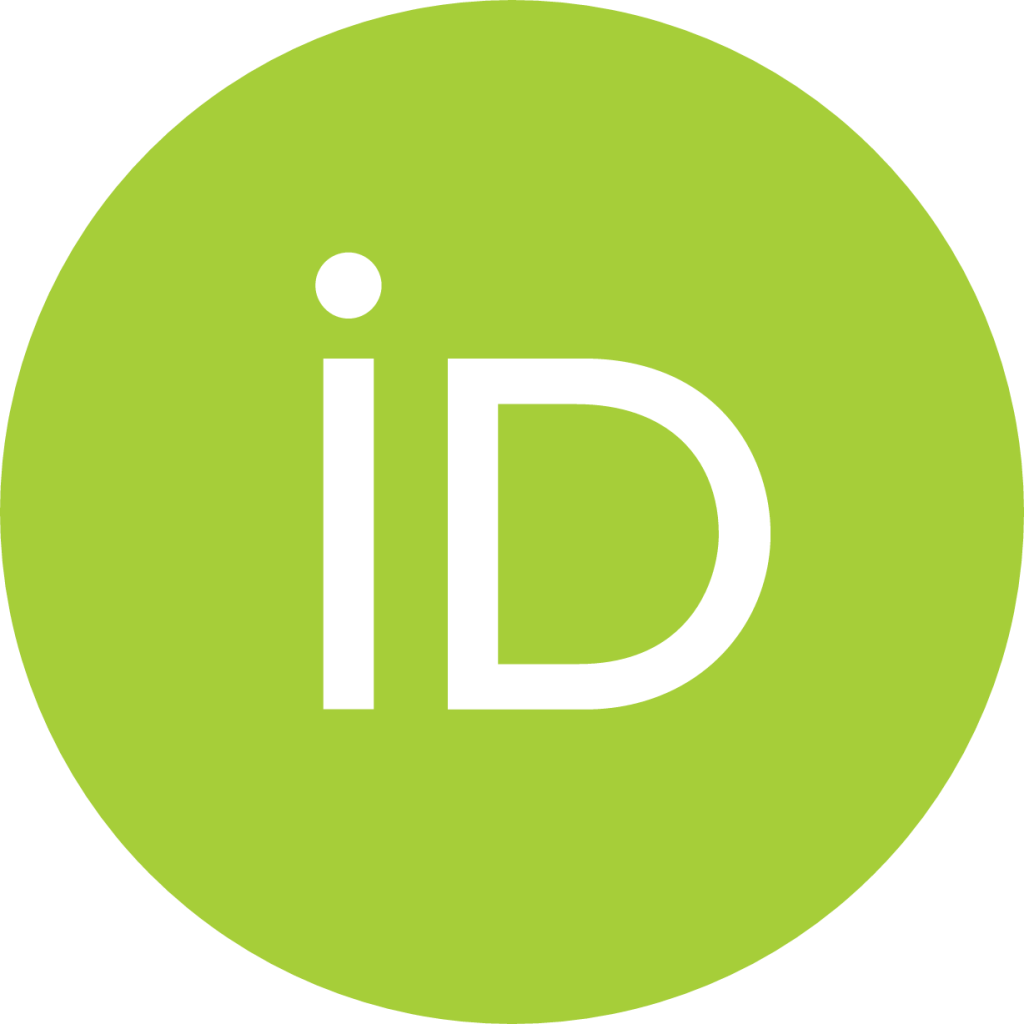}}}
\email{tamalmukhopadhyay7@gmail.com}
\affiliation{Department of Physics, Sister Nivedita University, DG Block(Newtown) 1/2, Action Area I, Kolkata-700156, India}

\author{Banadipa Chakraborty~\href{https://orcid.org/0009-0008-2378-0544}{\includegraphics[height=1em]{ORCID_iD_svg.png}}}
\email{jhelum.chakraborty@gmail.com}
\affiliation{Department of Physics, Sister Nivedita University, DG Block(Newtown) 1/2, Action Area I, Kolkata-700156, India}

\author{Anamika Kotal~\href{https://orcid.org/0009-0009-8426-3206}{\includegraphics[height=1em]{ORCID_iD_svg.png}}}
\email{kotalanamika31@gmail.com}
\affiliation{Department of Mathematics, Indian Institute of
Engineering Science and Technology,\\ Shibpur, Howrah-711103,
India.}

\author{Ujjal Debnath~\href{https://orcid.org/0000-0002-2124-8908}{\includegraphics[height=1em]{ORCID_iD_svg.png}}}
\email{ujjaldebnath@gmail.com}
\affiliation{Department of Mathematics, Indian Institute of
Engineering Science and Technology,\\ Shibpur, Howrah-711103,
India.}
\date{\today}

\begin{abstract}
   \singlespacing
     
     In this research, we have reconstructed the extended $f(\mathcal{P})$ cubic gravity and symmetric $f(\mathcal{Q})$ teleparallel gravity from the $(m,n)$-type Barrow Holographic Dark Energy (BHDE) model. We have derived the unknown functions $f(\mathcal{P})$ and $f(\mathcal{Q})$ in terms of $\mathcal{P}$ and $\mathcal{Q}$, assuming a flat, homogeneous, and isotropic universe. To constrain our model parameters, we employed cosmic chronometer datasets and Baryon Acoustic Oscillation datasets, utilizing Markov Chain Monte Carlo (MCMC) method. We analysed the behaviour and stability of each model throughout the universe's evolution by studying crucial parameters such as the deceleration parameter, equation of state (EoS) parameter $\omega_{DE}$, density parameter $\Omega(z)$ and the square of the speed of sound $v_s^2$. Additionally, we explored the cosmographic behaviour by plotting the jerk parameter, snap parameter, and lerk parameter against the redshift. Furthermore, we examined the $\omega'_{DE}-\omega_{DE}$ phase plane, the $(r,s^*)$, $(r,q)$ statefinder parameters, and the $Om(z)$ parameter offers profound revelations about the dynamics of the universe and the distinctive features of dark energy. Our analyses indicated that our model could produce a universe undergoing accelerated expansion with quintessence-type dark energy. These findings contribute to our understanding of the nature of dark energy and the evolution of the cosmos.
\end{abstract}

\maketitle

\section{Introduction}
In the late 20\textsuperscript{th} century, observational evidences from type Ia Supernova \cite{riess1998observational, Perlmutter_1998,perlmutter1999measurements} and large scale structure of the universe \cite{colless20012df, PhysRevD.69.103501} indicate that our universe is expanding in an accelerated manner which starts very recently \cite{shapiro2006we}. Additionally, the Cosmic Microwave Background Radiation (CMBR) anisotropies observed by the WMAP and Planck satellite \cite{komatsu2009five,ade2014planck} and Baryon Acoustic Oscillation \cite{cole20052df,percival2007measuring} provide crucial evidence supporting the claim. Though General Relativity is an outstanding theory, it fails to answer this late-time accelerated expansion. The concordance model of cosmology ($\Lambda$CDM) where $\Lambda$ is taken as so-called dark energy with negative pressure and positive energy density, has some serious drawbacks like ``Cosmological Coincidence Problem'' \cite{velten2014aspects,yoo2012theoretical} and others that remain unanswered \cite{weinberg2001cosmological}. Also, the Standard Model of Particle Physics cannot explain the existence of CDM (Cold Dark Matter) particles. To overcome this problem, several dynamical dark energy theories have been proposed. This kind of theory mainly contains a new degree of freedom as a scalar field with a Lagrangian where the potential or the kinetic term drives the accelerated expansion. The quintessence model \cite{peebles1988cosmology, Caldwell_1998,zlatev1999quintessence,carroll1998quintessence} is one of that kind which is mostly favored by physicists to be a potential candidate for dark energy. Also, other dark energy models with dynamic vacuum energy such as tachyon field \cite{Sen_2002}, k-essence \cite{armendariz2001essentials}, H-essence \cite{Wei_2005}, dilaton \cite{Gasperini_2001}, phantom \cite{Caldwell_2002}, Chaplygin gas \cite{GORINI_2006}, DBI-quintessence \cite{Gumjudpai_2009}, DBI-essence \cite{martin2008dbi} has been introduced to address the same problem.
\par Another approach is to add a modification to the geometric component of the Einstein-Hilbert action so that it can mimic the observed accelerated expansion of the universe without any ad-hoc dark energy term (See reviews \cite{clifton2012modified, nojiri2011unified, nojiri2017modified}). The concept of altering the traditional theory of gravity was initially explored within the framework of a specific class of models known as $f(R)$ gravity, where the Einstein-Hilbert action in General Relativity is generalised by introducing a function of the Ricci scalar curvature \cite{sotiriou2010f}. Inspired by this, several models on modified gravity are introduced and covered in \cite{Rastkar_2011, Nojiri_2005,ferraro2011non, PhysRevD.84.024020, Jamil_2012, Bamba_2010, Myrzakulov_2012}. In the past decade, two interesting modified gravity models have been proposed. One of them is Einstein Cubic Gravity(ECG), in which a 3\textsuperscript{rd} order contraction of the Riemann tensor is used \cite{PhysRevD.94.104005}. In their work, a specialised linearization technique is discussed to derive the ECG (Einstein cubic gravity) by incorporating a specific non-topological form of the term $\mathcal{P}$. This approach was pivotal in laying down the foundational principles of the ECG theory. Several works have been done on the ECG \cite{PhysRevD.97.064041, PhysRevD.99.024035, ECG3}. Later, extending the formulations of ECG, extended $f(\mathcal{P)}$ cubic gravity was discussed by Erices et al. \cite{PhysRevD.99.123527}. In this work, they added a function of non-topological term $\mathcal{P}$, which contains cubic contraction of specific Riemann tensor, to the Einstein-Hilbert action. The second one is the Teleparallel gravity, where the affine connection is considered flat, and the scalar quantity torsion $T$ is taken to be the mediator of gravity instead of the levi-civita curvature of GR \cite{saridakis2021modified,maluf2013teleparallel}. Based on this, several modifications to teleparallel gravity have been proposed: $f(T)$ gravity \cite{ferraro2007modified}, $f(T,B)$ gravity \cite{bahamonde2015modified} etc. Another interesting teleparallel gravity model is $f(\mathcal{Q})$ gravity, where the connection is symmetric, flat, and torsion-free but not metric compatible \cite{jimenez2018coincident}. The non-metricity scalar $\mathcal{Q}$ is used as the gravity mediator. The complete review of $f(\mathcal{Q})$ gravity is given in \cite{heisenberg2023reviewfqgravity}. Important works on $f(\mathcal{Q})$ gravity are given in \cite{capozziello2024preservingquantuminformationfq, capozziello2024gravitationalwavesfqnonmetric, PhysRevD.102.024057, f(Q)UjjalDebnath,PhysRevD.98.084043,PhysRevD.103.124001,Dimakis_2021,Jawadf(Q)}.
\par In cosmology, we generally solve the field equations to obtain the geometry of space-time as the solution. In the reconstruction method we usually employ the underlying geometry from observational or theoretical perspective and obtain the model equations which is consistent with the observation \cite{PhysRevD.74.086005, nojiri2006dark, PhysRevD.77.106005, nojiri2024well, GADBAIL2022137509}. In recent times, theoretical physicists have shown an increasing curiosity in investigating the potential connections between different models of Holographic dark energy (HDE) and various theories of modified gravity. This area of research has garnered significant attention as scientists seek to unravel the mysteries surrounding the nature of dark energy and its implications for our understanding of the universe \cite{chattopadhyay2013reconstruction, karami2011reconstructing,pasqua2013reconstruction,salako2015holographic, farooq2013reconstruction,saha2016reconstructions,saha2022reconstructing,ghosh2021reconstructions,debnath2015reconstructing, Gupta_f(Q)}.  In 2011, Karami et al. \cite{karami2011reconstructing} reconstructed the $f(R)$ gravity by taking into account two distinct approaches: the conventional models of holographic and new agegraphic dark energy, as well as the entropy-corrected versions of these models and showed that the dark energy EoS changes from quintessence to phantom regime in agreement with the observations. Chattopadhyay and Pasqua et al. have investigated the reconstruction of $f(T)$ and $f(R,T)$ gravity models from holographic dark energy and holographic Ricci dark energy \cite{chattopadhyay2013reconstruction, pasqua2013reconstruction}. They have found an action of $f(T)$, which showed consistency with holographic dark energy, and the reconstructed $f(R,T)$ attains classical stability under perturbation. Debnath \cite{debnath2015reconstructing} has explored the reconstruction of $f(G)$, $f(R)$, Einstein-Aether, $f(T)$ gravities based on entropy corrected formulation of $(m,n)$-type pilgrim dark energy. In a subsequent study, the researchers Ghosh, Debnath, and Chakraborty [1] explored the extended cubic gravity model represented by the function $f(\mathcal{P})$. They derived this model from two different approaches: the ordinary and entropy-corrected $(m,n)$-type holographic dark energy models, as well as the pilgrim dark energy model. Through their analysis, they investigated various cosmological parameters and examined the stability of the reconstructed model~\cite{ghosh2021reconstructions}. In \cite{koussour2023barrow}, the authors have worked on the Barrow holographic dark energy considering an anisotropic Bianchi type-I universe taking the geometrical background as symmetric $f(\mathcal{Q})$ teleparallel gravity. The Holographic principle was introduced by Hooft \cite{hooft1993dimensional} from Black hole thermodynamics \cite{bekenstein1973black,hawking1975particle}, which states that the information enclosed within a given volume of space can be encoded and represented on its boundary, analogous to how a hologram captures three-dimensional content on a two-dimensional surface. Many physicists believe that the Holographic principle plays a fundamental role in the quantum description of gravity. The applications of Holographic dark energy are given in \cite{li2004model,li2009probing,myung2011entropic,nojiri2006unifying,nojiri2019holographic,nojiri2020unifying}. Recently, newly proposed Barrow entropy correction for black hole gained much interest \cite{Barrow_2020}. Saridakis constructed Barrow holographic dark energy from the Barrow entropy \cite{PhysRevD.102.123525} and it shows some promising results over the general one \cite{anagnostopoulos2020observational,adhikary2021barrow,srivastava2021barrow}. The motivation of this paper is as follows: the definition of Barrow entropy makes it a more general description of nature as it comes from simple physical consideration, which resembles Tsallis nonextensive entropy \cite{TsallisEntropy, Wilk_2000, Tsallis_2013} and reduces to normal Bekenstein-Hawking entropy on a specific limit. The quantum-gravitational deformation correction to the horizon could possibly be the real description of nature and thus can solve the year-long questions of cosmology. The paper in Ref.~\cite{PhysRevD.102.123525} showed that the BHDE effectively encompasses the thermal evolution of the universe during epochs dominated by both dark matter and dark energy. Recently, a study showed that Barrow Entropic Dark Energy is a subset of a more general Holographic Dark Energy with a suitable IR cut-off \cite{nojiri2022barrow}. All these facts inspire us to do a study on the reconstruction of the functional form of $f(\mathcal{P})$ and $f(\mathcal{Q})$ from Barrow holographic dark energy as background evolution by taking the characteristic length in a specific form and then analyse its cosmological implications. In our study, we employed observational evidence from two distinct sources. Firstly, we d 31 direct measurements of the Hubble parameter obtained from Cosmic Chronometers (CC) data sets. Secondly, we incorporated 26 data points derived from Baryon Acoustic Oscillation (BAO) observations. By leveraging these empirical data points, we aimed to impose constraints on the parameters of our model. This allows us to determine the observational bounds of the parameters and, hence, to understand the undermining physical scenarios in depth. To achieve this, we employed the covariance matrix technique, which enabled us to minimise the associated errors and determine the optimal values for several key parameters within our proposed model. This paper is structured as follows: in Sect.~\ref{Sect:(m,n)typeBHDE} we have presented the $(m.n)$-type Barrow holographic dark energy, in Sect.~\ref{Sect:Cosmographic_parameters} we have discussed cosmographic parameters and different diagnostic tools in cosmology, in Sect.~\ref{Subsect:f(P)relations} and Sect.~\ref{Sect:f(Q)_relations} we have discussed basic relations from extended $f(\mathcal{P})$ cubic and symmetric $f(\mathcal{Q})$ teleparallel gravities, Sect.~\ref{Sect:f(P)_reconstruct} and Sect.~\ref{Sect:f(Q)_reconstruct} is dedicated for the reconstruction of functions $f(\mathcal{P})$, $f(\mathcal{Q})$ which are unknown. In Sect.~\ref{Sect:Methodology} we provide our methodology of parameter constraining using observational datasets, in Sect.~\ref{Sect:Result} we present our findings from the analysis and in the concluding section Sect.~\ref{Sect:Conclusion} we discuss and interpret the results we obtain from the analysis.

\section{\boldmath \texorpdfstring{$(m,n)$}{Lg}-type Barrow Holographic Dark Energy}\label{Sect:(m,n)typeBHDE}

Here, we set up the formulation of $(m,n)$ -type Barrow Holographic Dark Energy. The application of the holographic principle at the cosmological scale implies that the entropy at the horizon of the universe (corresponding to the largest distance) is directly proportional to the surface area of the horizon \cite{bousso2002holographic, hooft1993dimensional, Susskind1994TheWA, fischler1998holography}. This is analogous to the Bekenstein-Hawking entropy for black holes. Literature regarding the application of the holographic principle to explain dark energy can be found on \cite{Cohen_1999, Enqvist_2005, Gong_2004, Pavo_n_2006, ZHANG_2005}. 
\par Recently, motivated by the intricate surface structure of the COVID-19 virus, Barrow coined a different kind of entropy formulation for black holes where the quantum-gravitational effect can produce complicated, fractal structures on the structure of black hole event horizon. This formulation results in finite volume, but an infinite (or finitely large) area, and thus, the usual entropy-area relation is being modified. The Barrow entropy can be given by \cite{Barrow_2020},
\begin{equation}\label{Barrowentropy}
    S_B=\left(\frac{A}{A_0}\right)^{(1+\frac{\Delta}{2})}
\end{equation}
Here, $A$ denotes the surface area of the horizon, $A_0$ denotes the Planck area, and the Barrow exponent $\Delta$ signifies the quantification of quantum-gravitational deformity present in the black hole surface with $\Delta=0$ gives the most simple structure with Bakenstein-Hawking entropy and $\Delta = 1$ indicates increasingly complex horizon structure.
To apply this entropy relation in the scenario of dark energy, we will recall that the usual holographic dark energy density is given as $\rho_{DE}L^4\leq S$ \cite{Wang_2017} and using this we can obtain the entropy area relation for Barrow holographic dark energy from \eqref{Barrowentropy} \cite{Saridakis_2020},
\begin{equation}\label{rho_BHDE}
    \rho_{BHDE}= CL^{\Delta-2} 
\end{equation}
Here, C denotes a parameter with the dimension of $[L]^{-\Delta-2}$ and $L$ is the length of the horizon. We can note that, at $\Delta=0$,  Eq.~\eqref{rho_BHDE} takes the form of usual holographic dark energy density $\rho_{DE}=3c^2M_P^2L^{-2}$ with $C=3c^2M_P^2$ where $M_P$ is the Planck mass and $c$ being a constant parameter (not the speed of light in vacuum). The presence of $\Delta$ deviates the Barrow holographic dark energy from the usual holographic dark energy and exhibits intriguing cosmological phenomena.
\par The horizon length $L$ is a very crucial quantity in the case of holographic dark energy as it defines the length at which the quantum gravity effects become relevant. The characteristic length also imposes certain conditions on the energy density of the proposed holographic dark energy model and hence asserts the cosmological behaviour of the model universe to a great extent. One of the most well-known connections between holography and gravity comes from the AdS/CFT correspondence, proposed by Juan Maldacena \cite{Maldacena_1999}. This duality suggests that a gravity theory in a higher-dimensional AdS space can be equivalent to a conformal field theory on the boundary of that space. Here, the holographic length plays a crucial role in linking the bulk (gravity) dynamics to the boundary. The holographic principle has led to the idea that gravity itself might be an emergent phenomenon, arising from more fundamental quantum degrees of freedom. Erik Verlinde's theory of ``emergent gravity'' suggests that gravity is not a fundamental force but emerges from the entropic force associated with changes in the information content on the holographic screen, which could be related to the holographic length \cite{Verlinde_2011}. The choice of the characteristic scale $L$ is the topic of debate over the last few years \cite{hovrava2000probable,thomas2002holography,hsu2004entropy, li2004model,cai2007dark,wei2008new,gao2009holographic, huang2012holographic}. Several pieces of literature proposed that $L$ is the future event horizon \cite{li2004model}, the conformal age of the universe \cite{wei2008new}, and the Ricci scalar of the universe \cite{gao2009holographic}. Recently, Ling and Pan introduced a new kind of generalised holographic scale defined by two numbers $(m,n)$ \cite{LING_2013}. Indeed, while lacking direct physical motivation, this characteristic scale offers greater theoretical flexibility, allowing for better alignment with observational data. 
\newline The work of our paper aims to draw a correspondence between $(m,n)$-type BHDE and $f(\mathcal{P})$, $f(\mathcal{Q})$ gravity and by doing so we want to investigate how the aforementioned modified gravity models can inherently adapt the predictions given by Barrow holographic dark energy and thus can be aligned with the quantum theory of gravity. The holographic principle might influence the form of these unknown functions, particularly in how they reconstruct from different energy conditions or dark energy models. In this regard, the characteristic length $L$ plays a very important role in our analysis. For the sake of the purpose that the dark energy will drive the accelerated expansion, the horizon length must be the future event horizon, and for the $(m,n)$-type BHDE, the IR cut-off (Horizon length) becomes the conformal age of the universe \cite{LING_2013, farooq2013reconstruction}.
Thus, the characteristic length is given by,
\begin{equation}\label{mnLength}
    L=\frac{1}{a^m(t)}\int_t^\infty a^n(t')dt'
\end{equation}
The numbers $(m,n)$ are real in nature, and they need not be integers for the phenomenological level.
Now, to obtain the expression of $L$, let’s proceed with the assumption: the scale factor of our universe follows a power law form given by,
\begin{equation}\label{a(t)value}
    a(t) = a_0 t^\delta,
\end{equation}
The term $a_0$ represents the scale factor’s value at the current epoch of our universe, and $\delta>1$ denotes a positive constant parameter. Evaluating \eqref{mnLength} we obtain,
\begin{equation}\label{Lcalculatedvalue}
     L=\frac{t(a_0t^\delta)^{(n-m)}}{1+\delta n}
\end{equation}
Now, we have to express the characteristic length $L$ of eq.~\eqref{Lcalculatedvalue} in terms of $\mathcal{P}$ and $\mathcal{Q}$. After that, we will put the calculated value of $L$ as a function of $\mathcal{P}$ and $\mathcal{Q}$ in the eq.~\eqref{rho_BHDE}. Then, we will equate the dark energy densities of $(m,n)$-type BHDE and $f(\mathcal{P})$, $f(\mathcal{Q})$ gravities to obtain the unknown functions. The above method is described in the section \ref{Sect:f(P)_reconstruct} and \ref{Sect:f(Q)_reconstruct}. This way, we can connect the $(m,n)$-type BHDE with the frameworks of $f(\mathcal{P})$ and $f(\mathcal{Q})$ gravity.

\section{Review of Cosmographic Parameters and Diagnostic tools}\label{Sect:Cosmographic_parameters}
Cosmographic parameters provide an independent way to describe the evolution of the universe. They are also useful tools to differentiate among several dark energy models \cite{wang2009probing}. These parameters can possibly be measured directly from the observational redshift data, and using this, one can probe the nature of the expansion history of the cosmos without making any prior assumptions. These results can directly be interpreted and used to test the feasibility of several cosmological models. The parameters are defined as a Taylor series expansion of Hubble parameter about redshift $z$ keeping the origin at $z=0$ \cite{MattVisser_2004}
\begin{equation}
\label{cosmographicparameters}
q\equiv-\frac{\ddot{a}}{H^2a},\ \ j\equiv\frac{\dddot{a}}{H^3a},
\ \ s\equiv\frac{\ddddot{a}}{H^4a} \ \ \mbox{and}\ \ l\equiv\frac{a^{(5)}}{H^5a}
\end{equation}
Here, $q$, $j$, $s$, and $l$ signify deceleration parameter, jerk parameter, snap parameter and lerk parameter respectively. The dots and numbers in the bracket indicate the time derivative. Here, $q$, the deceleration parameter, indicates the expansion rate of the universe. For $q$ being positive, the universe is expanding in a decelerated manner; for negative $q$, the universe is expanding with acceleration, and for $q=0$, the rate of expansion of the universe is constant. The jerk parameter $j$ signifies if the accelerated expansion of the universe is increasing ($j>0$), decreasing ($j<0$), or constant ($j=0$) \cite{barboza2017constraints}. In a recent paper, Caldwell et al. argued that the jerk parameter could be the first non-CMB cosmological test that can probe the spatial curvature of the universe \cite{caldwell2004expansion}. Similarly, the snap parameter and lerk parameter measure the rate of change of jerk and snap parameters, respectively. The rate of change is increasing for positive $(>0)$ value, constant for $(=0)$ and decreasing for negative $(<0)$ value. The recent works of cosmographic parameters regarding the model independent reconstructions of modified gravity models can be found in \cite{capozziello2019extended, CAPOZZIELLO2022137229, CAPOZZIELLO2023101346}.
\par Now, from the Eq. \eqref{cosmographicparameters} the cosmographic parameters can be expressed by,
\begin{equation}\label{q,j,l,s_final}
    \begin{split}
         q &= -1-\frac{\dot{H}}{H^2}\\
    j &= \frac{\ddot{H}}{H^3}-3q-2 \\
     &= q(2q+1)+(1+z)\frac{dq}{dz}\\ 
    s &= \frac{H^{(3)}}{H^4}+4j+3q(q+4)+6\\
     &= -(1+z)\frac{dj}{dz}-j(2+3q)\\ 
    l &= \frac{H^{(4)}}{H^5}-24-6q-3q^2-10j(q+2)+5s\\
     &= -(1+z)\frac{ds}{dz}-s(3+4q) 
    \end{split}
\end{equation}
The Equation of state (EoS) parameter for dark energy can be defined as,
\begin{equation}\label{omega_DE}
    \omega_{DE} = \dfrac{p_{DE}}{\rho_{DE}} = -1 + \frac{1+z}{3} \cdot \frac{2 \left[1 - \Omega_m(z)\right]}{H(z)^2} \cdot \frac{d H(z)^2}{dz}
\end{equation}
Where, $\Omega_m(z)$ is the matter-density parameter at redshift $z$ and given by,
\[\Omega_m(z) = \frac{\Omega_{m0} (1+z)^3}{H(z)^2/H_0^2}\] Here, $\Omega_{m0}$ is the density parameter of matter in today's universe.
The table.~\ref{tab:omegarange} shows physical significance of EoS parameter and its bounds.
\begin{table}[htbp] 
\centering
\renewcommand{\arraystretch}{1.5}
\begin{tabular}{|p{0.4\linewidth}|p{0.4\linewidth}|}
\hline
\textbf{Dark Energy Type} & \textbf{EoS Parameter Range} \\ \hline
Quintom Model & $\omega_{DE} > -1$, crossing the $\omega_{DE} = -1$ line \\ \hline
Quintessence & $-1 < \omega_{DE} < -\frac{1}{3}$ \\ \hline
Phantom Model & $\omega_{DE} < -1$ \\ \hline
\end{tabular}
\caption{Classification of dark energy Models using EoS Parameter}
\label{tab:omegarange}
\end{table}
Another crucial quantity for analyzing the stability of cosmological models is the square of the speed of sound, which can be expressed as \cite{kim2008instability},
\begin{equation}\label{v_s^2}
    v_s^2 = \dfrac{dp_{DE}}{d\rho_{DE}} = v^2(z) = \omega_{DE}(z) + \frac{(1+z)}{3} \frac{d \omega_{DE}(z)}{dz}
\end{equation}
The evolution of the model is classically stable if $0<v_s^2<1$, otherwise unstable for a given perturbation.
\par In $2005$ Caldwell and Linder \cite{PhysRevLett.95.141301} proposed another kind of diagnostic tool called $\omega'-\omega$ phase plane to differentiate between dark energy models where $\omega'_{DE}$ is given by,
\begin{equation}\label{omega_prime}
    \omega'_{DE} = \dfrac{d}{d\chi} \omega_{DE} 
\end{equation}
Here, the variable $\chi = \ln{a(t)}$. This phase plane has two regions based on the following conditions:
\begin{table}[htbp]
\centering
\renewcommand{\arraystretch}{1.5}
\begin{tabular}{|p{0.3\linewidth}|p{0.3\linewidth}|p{0.3\linewidth}|}
\hline
\textbf{Condition} & \textbf{Name} & \textbf{Remarks} \\ \hline
$\omega'_{DE} > 0, \: \omega_{DE} < 0$ & Thawing Region & Less accelerated expansion. \\ \hline
$\omega'_{DE} < 0, \: \omega_{DE} < 0$ & Freezing Region & More accelerated expansion. \\ \hline
\end{tabular}
\caption{Specifications of $\omega'_{DE}-\omega_{DE}$ plot}
\label{tab:omega'-omega}
\end{table}
\par In $2003$ Sahni et al. \cite{sahni2003statefinder} proposed a new geometrical diagnostic tool known as the statefinder pair $(r,s^*)$ to probe the characteristics of dark energy without relying on a specific model. The statefinder parameter $(r,s^*)$ depends on the third derivative of the Hubble parameter with respect to time and can be expressed by,
\begin{equation} \label{r_s_parameter}
r = \frac{\stackrel{...}{a}}{a H^3}, ~~s^* = \frac{r - 1}{3\,(q - 1/2)}.
\end{equation}
Here, $q$ denotes the deceleration parameter, and by the definition given in \eqref{cosmographicparameters}, we can identify $r$ with the usual jerk parameter $j$. In the past few years, several works have been done using statefinder parameters to classify different dark energy models and can be found on \cite{chakraborty2012statefinder,panotopoulos2008statefinder,solanki2022statefinder, Alvarez_2022, Chang:2007jr,evans2005geometrical,alam2003exploring,feng2008statefinder}. The graphical analysis of statefinder parameters can be interpreted by following ways given in Table.~\ref{tab:r_s_parameter}.
\begin{table}[htbp]
\centering
\renewcommand{\arraystretch}{2.0}
\begin{tabular}{|p{0.3\linewidth}|p{0.3\linewidth}|p{0.3\linewidth}|}
\hline
\textbf{Value} & \textbf{Type} & \textbf{Description} \\ \hline
$(r < 1, s^* > 0)$ & Quintessence & Gives quintessence type dark energy. \\ \hline
$(r = 1, s^* = 0)$ & ΛCDM & Gives ΛCDM like dark energy. \\ \hline
$(r > 1, s^* < 0)$ & Chaplygin Gas (phantom) &  Gives an alternative regime where dark energy is described as a perfect fluid, characterised by an EoS derived from the equation discovered by Chaplygin \cite{chaplygin1944gas}. \\ \hline
\end{tabular}
\caption{Classification of DE models based on $(r,s^*)$ parameters}
\label{tab:r_s_parameter}
\end{table}

The \( \text{Om}(z) \) diagnostic tool is given by the formula \cite{sahni2008two}:

\begin{equation}\label{Om(z)_parameter}
\text{Om}(z) = \frac{E(z)^2 - 1}{(1 + z)^3 - 1}
\end{equation}
where \( E(z) \) is the dimensionless Hubble parameter, defined as:
\[E(z) = \frac{H(z)}{H_0}\]
Here, \( H(z) \) represents Hubble parameter at redshift \( z \), and \( H_0 \) represents Hubble constant.
\begin{itemize}
   \item \textbf{Scenario 1: \( \text{Om}(z) > \Omega_m \)}

    When \( \text{Om}(z) \) is greater than the matter density parameter \( \Omega_m \), this indicates a dark energy model with a quintessence-like equation of state, where \( w(z) < -1 \). This scenario suggests a universe where dark energy is more dominant, causing a faster acceleration of cosmic expansion.

    \item \textbf{Scenario 2: \( \text{Om}(z) < \Omega_m \)}

    If \( \text{Om}(z) \) is less than \( \Omega_m \), it implies a phantom-like dark energy model, where the equation of state parameter \( w(z) > -1 \). This indicates a universe with a less dominant dark energy component, leading to slower acceleration of the cosmic expansion.

    \item \textbf{Scenario 3: \( \text{Om}(z) = \Omega_m \)}

    In the case where \( \text{Om}(z) = \Omega_m \), the universe behaves according to the standard \(\Lambda\)CDM model, where dark energy is characterised by a cosmological constant (\( w(z) = -1 \)). This results in a constant \( \text{Om}(z) \), consistent with a non-evolving dark energy component.

\end{itemize}

\section{Reconstruction of Extended cubic gravity model \texorpdfstring{$f(\mathcal{P})$}{Lg}}\label{Sect:f(P)gravity}
\subsection{\large Basic relations of \texorpdfstring{$f(\mathcal{P})$}{Lg} gravity}\label{Subsect:f(P)relations}
In the following segment, we will set up the primary formulations of $f(\mathcal{P})$ gravity within the context of a flat, homogeneous, and isotropic FLRW universe, which is described as,
\begin{equation}\label{FLRWmetric}
    ds^2 = -dt^2+a^2(t)[dr^2+r^2(d\theta^2 + \sin^2\theta d\phi)]
\end{equation}
Where $a(t)$ denotes the scale factor of our universe. The $f(\mathcal{P})$ cubic gravity action is just a modification to the Einstein-Hilbert action for GR by a function $f(\mathcal{P})$, where $\mathcal{P}$ is a non-topological cubic invariant is written as \cite{PhysRevD.64.024028,PhysRevD.99.123527}:
\begin{equation}\label{f(P)action}
   S= \int \sqrt{-g}d^4x \left[\frac{R}{2\kappa}+f(\mathcal{P})\right] 
\end{equation}
Here, $\kappa = 8\pi G$ and $P$ can be written as \cite{bueno2016einsteinian},
\begin{equation}\label{P}
\begin{split}
\mathcal{P}=&\beta_1 R_{\mu\nu}^{\rho\sigma}R_{\rho\sigma}^{\gamma\delta}R_{\gamma\delta}^{\mu\nu}+\beta_2 R_{\mu\nu}^{\rho\sigma}R_{\rho\sigma}^{\gamma\delta}R_{\gamma\delta}^{\mu\nu} +\beta_3R^{\sigma\gamma}R_{\mu\nu\rho\sigma}R_{\gamma}^{\mu\nu\rho}\\
&+\beta_4 R R_{\mu\nu\rho\sigma}R^{\mu\nu\rho\sigma}+\beta_5 R_{\mu\nu\rho\sigma}R^{\mu\rho}R^{\nu\sigma} \\
&+\beta_6 R_{\mu}^{\nu}R_{\nu}^{\rho}R_{\rho}^{\mu}+\beta_7 R_{\mu\nu}R^{\mu\nu}R+\beta_8 R^3
\end{split}
\end{equation}
Where $\beta_1, \ \beta_2, \ \beta_3, \ \beta_4, \ \beta_5, \ \beta_6, \ \beta_7, \ \beta_8$ are some constant parameters. Interestingly, the non-topological cubic term $\mathcal{P}$ arises from the 3\textsuperscript{rd} order contraction of the Riemann tensor. This geometric condition subsequently gives the second-order field equations that satisfy certain constraints involving the above parameters. \cite{PhysRevD.101.103534}
\begin{equation}\label{constrain1}
\beta_7=\frac{1}{12}\big[3\beta_1-24\beta_2-16\beta_3-48\beta_4-5\beta_5-9\beta_6\big],
\end{equation}
\begin{equation}
\beta_8=\frac{1}{72}\big[-6\beta_1+36\beta_2+22\beta_3+64\beta_4+5\beta_5+9\beta_6\big],
\end{equation}
\begin{equation}\label{constrain6}
\beta_6=4\beta_2+2\beta_3+8\beta_4+\beta_5
\end{equation}
In addition, we can define another parameter as \cite{ASardar_Debnath, PhysRevD.101.103534}
\begin{equation}
\beta=(-\beta_1+4\beta_2+2\beta_3+8\beta_4)
\end{equation}
Now, using these constraint relations \eqref{constrain1}-\eqref{constrain6} the cubic invariant $\mathcal{P}$ takes the form in the context of FLRW metric \eqref{FLRWmetric} as,
\begin{equation}\label{Pvalue}
 \mathcal{P}=6\beta H^4 (2H^2+3\dot{H}),
\end{equation}
Which contains the second-order derivative of the scale factor $a(t)$,  leading to the field equations of the second order.
Hence, we can readily obtain the corresponding Friedmann equations by varying the action given in \eqref{f(P)action} and taking it as zero. The modified field equations take the form as given by \cite{PhysRevD.99.123527, ASardar_Debnath},

\begin{equation}\label{friedmann1}
3H^2=\kappa(\rho_m+\rho_{f(P)}),
\end{equation}
\begin{equation}\label{friedmann2}
3H^2+2\dot{H}=-\kappa(p_m+p_{f(P)})
\end{equation}

Here, $H =\frac{\dot{a}}{a}$ is the Hubble parameter. For the mathematical ease, we will use $\kappa = 1$ hereafter. The dark energy density and pressure for $f(\mathcal{P})$ gravity can be written as,

\begin{equation}\label{rhof(P)}
   \rho_{f(P)}=-f(\mathcal{P})-18\beta H^4\left[H\partial_t-H^2-\dot{H}\right]f'(\mathcal{P}) 
 \end{equation}
 and,
 \begin{equation}\label{pf(P)}
 p_{f(P)}= f(\mathcal{P})+6\beta H^3\left[H\partial_t^2+2(H^2+\dot{H})\partial_t-3H^3-5H\dot{H}\right]f'(\mathcal{P})    
 \end{equation}
 Where $\partial_t$ and $\partial_t^2$ have their usual meaning. Dots and prime denote the derivative with respect to cosmic time and the non-topological term $\mathcal{P}$, respectively.
 \newline
Now, the fractional density parameters can be defined as,
 \begin{equation}\label{Omegam}
\Omega_m=\frac{\rho_{m}}{3H^2},
\end{equation}
\begin{equation}\label{OmegaDE}
\Omega_{DE}=\frac{\rho_{DE}}{3H^2}
\end{equation}
Which yields the constraint relation followed from \eqref{friedmann1} as,
\begin{equation}\label{modifiedfriedman1}
\Omega_m+\Omega_{DE}=1
\end{equation}
The matter component and the dark energy components adhere to their respective continuity equations as the matter part is not coupled with the geometry, and it could be given by,
\begin{equation}\label{mattercontinuity}
\dot{\rho}_{m}+3H(\rho_{m}+p_{m})=0.
\end{equation}
and,
\begin{equation}\label{DEcontinuity}
\dot{\rho}_{DE}+3H(\rho_{DE}+p_{DE})=0
\end{equation}
The solution of the \eqref{mattercontinuity} is given by,
\begin{equation}\label{rho_mvalue}
    \rho_{m} = \rho_{m0} (1 + z)^{3(1 + \omega_{m})},
\end{equation}
Here, $\rho_{m0}$ signifies the dark matter density in today's universe. Again, from the definition $\omega_m = \dfrac{p_m}{\rho_m}$ and this follows,
\begin{equation}\label{p_mvalue}
    p_{m} = \omega_{m} \rho_{m0} (1 + z)^{3(1 + \omega_{m})}
\end{equation}
Here, $\omega_m$ is the constant EoS parameter for dark matter. The interrelation between the scale factor $a(t)$ and the redshift parameter $z$ can be written as,
\begin{equation}\label{redshift_a(t)}
    1+z = \frac{1}{a(t)}
\end{equation}

\par Now, by using \eqref{a(t)value}, we derive certain relationships based on the definition of the Hubble parameter,
\begin{align}
    H &=\frac{\delta}{t} \label{H_delta}\\
    \dot{H} &=-\frac{\delta}{t^2} \label{H_dot} \\
    \ddot{H} &=\frac{2\delta}{t^3} \label{H_ddot}
\end{align}

\subsection{\large Reconstruction of \texorpdfstring{$f(\mathcal{P})$}{Lg} and dynamical analysis}\label{Sect:f(P)_reconstruct}
In this segment, we have deduced the functional expression for the unknown function $f(P)$. Using the relations \eqref{H_delta}, \eqref{H_dot} we have obtained $P$ from Eq. \eqref{Pvalue} as,
\begin{equation}\label{P_deltavalue}
    \mathcal{P}=6\beta \frac{\delta^5 (2\delta-3)}{t^6}
\end{equation}
Now, the characteristic length $L$ can be written using Eqs. \eqref{H_delta}, \eqref{H_dot}, \eqref{a(t)value}, \eqref{P_deltavalue} and \eqref{Lcalculatedvalue} in terms of $\mathcal{P}$ as,
\begin{equation}\label{LintermsofM}
     L=\dfrac{M_1}{\mathcal{P}^\frac{2\zeta}{3}}
\end{equation}
Where, the values of $M_1$ and $\zeta$ are given by,  
 $M_1=\frac{a_0^{(n-m)}}{|1+\delta n|} 
 [6\beta\delta^5 (2\delta-3)]^{\frac{2\zeta}{3}}$,   $\zeta=\frac{\delta (n-m)+1}{4}$. The term $n\delta + 1>0$ is associated with the expansion of the particle horizon, and $n\delta +1<0$ is associated with the contraction of the future event horizon.
 Then, we have calculated $\rho_{f(P)}$ using the values of $H, \dot{H}$ and $P$ from \eqref{H_delta}, \eqref{H_dot}, \eqref{P_deltavalue}:
 \begin{equation}\label{rho_fp_calculated}
     \rho_{f(P)}=-f(\mathcal{P})+\frac{18f''(\mathcal{P})\mathcal{P}^2}{(2\delta-3)}+\frac{3(\delta-1)\mathcal{P}f'(\mathcal{P})}{(2\delta-3)}
 \end{equation}
 Here, primes indicate the derivative with respect to $\mathcal{P}$.
\newline Now, to construct extended cubic gravity $f(\mathcal{P})$ from Barrow holographic dark energy, we have to make an analogy by equating the dark energy densities given by \eqref{rho_fp_calculated} and \eqref{rho_BHDE}:
 \begin{equation}\label{f_P_Equation}
      18f''(\mathcal{P})\mathcal{P}^2+3(\delta-1)\mathcal{P}f'(\mathcal{P})+(3-2\delta)f(\mathcal{P})=C(2\delta-3)\left[\frac{M_1}{\mathcal{P}^\frac{2\zeta}{3}}\right]^{\Delta-2}
 \end{equation}
 Where we have put $ L=\dfrac{M_1}{\mathcal{P}^\frac{2\zeta}{3}}$ in the equation~\eqref{rho_BHDE}. Integrating this, we get,
 \begin{equation}\label{f_P_unsimplified}
     f(\mathcal{P}) = \frac{C_1\mathcal{P}^{X_1+Y_1}M_1^\Delta \mathcal{P}^{-\frac{2\Delta \zeta}{3}}}{D_3M_1^2D_1}+\frac{C_1\mathcal{P}^{X_2+Y_2}M_1^\Delta \mathcal{P}^{-\frac{2\Delta \zeta}{3}}}{D_3M_1^2D_2}+C_2 \mathcal{P}^{X_1}+C_3 \mathcal{P}^{X_2}
 \end{equation}
 By further simplifications, finally, we have arrived at
\begin{equation}\label{f_P_Expression}
     \boxed{f(P)=D_4M_1^{\Delta-2}\mathcal{P}^{X_1+Y_1-\frac{2\Delta \zeta}{3}} + D_5M_1^{\Delta-2}\mathcal{P}^{X_2+Y_2-\frac{2\Delta \zeta}{3}}+C_2\mathcal{P}^{X_1}+C_3\mathcal{P}^{X_2}} 
 \end{equation}
Here, $C_2$ and $C_3$ are arbitrary constants. The value of other parameters are given by,  
\begin{equation}
\begin{split}
    C_1 &=4C\sqrt{3-2\delta} \\
    X_1 &=\frac{\sqrt{3 - 2\delta}\left(-\frac{7\sqrt{2}\sqrt{3 - 2\delta} - \sqrt{2}\sqrt{3 - 2\delta}\delta}{2(-3 + 2\delta)} - \frac{1}{2}\sqrt{-16 + \frac{98(3 - 2\delta)}{(-3 + 2\delta)^2} - \frac{28(3 - 2\delta)\delta}{(-3 + 2\delta)^2} + \frac{2(3 - 2\delta)\delta^2}{(-3 + 2\delta)^2}}\right)}{6\sqrt{2}} \\
    Y_1 &= \frac{1}{12} \left(-7 + \delta + \sqrt{3 - 2\delta}\sqrt{\frac{25 + 2\delta + \delta^2}{3 - 2\delta}} + 16\zeta\right) \\
    D_1 &= -7 + \delta + \sqrt{3 - 2\delta}\sqrt{\frac{25 + 2\delta + \delta^2}{3 - 2\delta}} - 8(-2 + \Delta)\zeta \\
    D_3 &= \sqrt{\frac{25 + 2\delta + \delta^2}{3 - 2\delta}}
\end{split}
\end{equation}
\begin{equation}
\begin{split}
    X_2 &= \frac{\sqrt{3 - 2\delta}\left(-\frac{7\sqrt{2}\sqrt{3 - 2\delta} - \sqrt{2}\sqrt{3 - 2\delta}\delta}{2(-3 + 2\delta)} + \frac{1}{2}\sqrt{-16 + \frac{98(3 - 2\delta)}{(-3 + 2\delta)^2} - \frac{28(3 - 2\delta)\delta}{(-3 + 2\delta)^2} + \frac{2(3 - 2\delta)\delta^2}{(-3 + 2\delta)^2}}\right)}{6\sqrt{2}} \\
    Y_2 &= \frac{1}{12} \left(-7 + \delta - \sqrt{3 - 2\delta}\sqrt{\frac{25 + 2\delta + \delta^2}{3 - 2\delta}} + 16\zeta\right)\\
    D_2 &= 7 - \delta + \sqrt{3 - 2\delta}\sqrt{\frac{25 + 2\delta + \delta^2}{3 - 2\delta}} + 8(-2 + \Delta)\zeta\\
    D_4 &= \frac{C_1}{D_3D_1}\\
    D_5 &= \frac{C_1}{D_3D_2}
\end{split}
\end{equation}

\begin{figure}[htbp]
    \centering
    \includegraphics[width=0.6\textwidth]{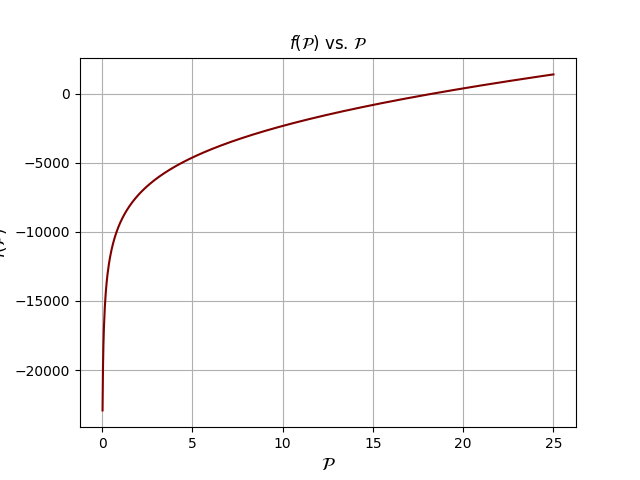}
    \caption{Plot of $f(\mathcal{P})$ against the cubic invariant $\mathcal{P}$ for the reconstructed extended cubic gravity $f(\mathcal{P})$ based on Barrow holographic dark energy.}
    \label{fig:f(P)vsPplot}
\end{figure}

It should be noted that the denominators of \eqref{f_P_unsimplified} and constants are $\neq 0$. Eq. \eqref{f_P_Expression} is the reconstructed functional form of $f(\mathcal{P})$ from $(m,n)$-type Barrow holographic dark energy. The graph depicted in Fig.~\ref{fig:f(P)vsPplot} illustrates the relationship between the function $f(\mathcal{P})$ and the cubic invariant $\mathcal{P}$. It is evident from the plot that as the cubic invariant $\mathcal{P}$ increases, the function $f(\mathcal{P})$ ascends accordingly.  

\section{Reconstruction of \texorpdfstring{$f(\mathcal{Q})$}{Lg} teleparallel gravity}\label{Sect:f(Q)_gravity}
\subsection{\large Basic relations of \texorpdfstring{$f(\mathcal{Q})$}{Lg} gravity}\label{Sect:f(Q)_relations}
All alternative theories of General Relativity mainly use two kinds of connections through which gravitational interaction is signified: (a) The Levi-Civita connection, which relies on the metric tensor's connection and its curvature (b) The Weitzenböck connection, based on tetrads of a metric, offering a curvature-free, metric-compatible connection with torsion. Another approach was introduced by Jim\'enez et al. \cite{jimenez2018coincident} where a symmetric teleparallel connection is used, which is curvature-free and torsion-free, but not metric-compatible. This kind of modified gravity model is called $f(\mathcal{Q})$ teleparallel gravity, where the force of gravitation is mediated through a scalar $\mathcal{Q}$, which exhibits a non-metricity nature. The reader is encouraged to see a more detailed review of metric-affine gravity in \cite{Capozziello_2022}. The corresponding action for the symmetric $f(\mathcal{Q})$ teleparallel gravity can be written as \cite{jimenez2018coincident,PhysRevD.102.024057,f(Q)UjjalDebnath},
\begin{equation}\label{f(Q)action}
    S = \int \sqrt{-g}d^4x\left[\frac{f(\mathcal{Q})}{2}+\mathcal{L}_m\right]
\end{equation}
Where the Lagrangian density $\mathcal{L}_m$ is associated with the matter content of the universe and the function $f(\mathcal{Q})$ is dependent on the scalar $Q$. In tensorial form $Q$ can be written as \cite{jimenez2018coincident}:
\begin{equation}\label{Qtensor}
    \mathcal{Q} = -\mathcal{Q}_{\gamma \mu \nu}V^{\gamma \mu \nu}
\end{equation}
Where, $V^{\gamma \mu \nu}$ is called superpotential and given by,
\begin{equation}\label{f(Q)superpotential}
    4V^{\gamma}_{\mu\nu} = -\mathcal{Q}^{\gamma}_{\mu\nu} + 2\mathcal{Q}_{({\mu}^{\gamma}\nu)} - \mathcal{Q}^{\gamma}g_{\mu\nu} - \tilde{\mathcal{Q}}^{\gamma}g_{\mu\nu} - \delta_{({\gamma}^Q{\nu})}
\end{equation}
Where the traces are given by,
\begin{eqnarray}\label{Qtrace}
    \mathcal{Q}_{\gamma} = \mathcal{Q}_{\gamma\:\mu}^{\mu} \ , \ 
    \tilde{\mathcal{Q}}_{\gamma} = \mathcal{Q}^{\mu}_{\:\:\gamma\mu}
\end{eqnarray}
Now, let us consider a universe is permeated with a perfect fluid, where the energy-momentum tensor is described by:
\begin{equation}\label{f(Q)energymomentum}
    T_{\mu\nu} = -\dfrac{2\delta(\sqrt{-g} \mathcal{L}_{m})}{\sqrt{-g}\delta g^{\mu\nu}}
\end{equation}
Then, the field equation takes the form,
\begin{equation}\label{f(Q)fieldequation}
    \dfrac{2}{\sqrt{-g}}\nabla_{\gamma}\left(\sqrt{-g}f_{\mathcal{Q}} V^{\gamma}_{\:\: \mu\nu}\right) + \dfrac{1}{2}g_{\mu\nu}f(\mathcal{Q}) + f_{\mathcal{Q}}\left(V_{\mu\gamma i}\mathcal{Q}_{\nu}^{\:\:\gamma i} - 2\mathcal{Q}_{\:\gamma i \mu}V^{\gamma i}_{\:\: \nu}\right) = -T_{\mu\nu} 
\end{equation}
Here, $f_{\mathcal{Q}} = \frac{df}{d\mathcal{Q}}$ and by varying the action in Eq.\eqref{f(Q)action} with respect to the connection, it yields
\begin{equation}\label{f(Q)actionvariationconnection}
    \nabla_{\mu}\nabla_{\gamma}\left(2\sqrt{-g}f_{\mathcal{Q}} V^{\gamma}_{\:\:\mu\nu}\right) = 0
\end{equation}
In our current research, we consider the universe to be spatially flat, with matter and energy distributed uniformly throughout, and exhibiting the same properties in all directions. The metric element of such a universe will be,
\begin{equation}\label{metricf(Q)}
    ds^2 = -dt^2 + a^2(t) \delta_{\mu\nu} dx^{\mu} dx^{\nu}
\end{equation}
Using this line element, we can express the trace of $Q$ can be written as,
\begin{equation}\label{Qvalue}
    \mathcal{Q} = 6H^2
\end{equation}
Now, the modified field equations are given as \cite{f(Q)UjjalDebnath},
\begin{eqnarray}\label{f(Q)friedmann}
    3H^2 &= \rho+\rho_q \\
    3H^2+2\dot{H} &= -(p+p_q)
\end{eqnarray}
Where, the energy density and pressure for the dark energy in $f(\mathcal{Q})$ gravity are given by,
\begin{eqnarray}
    \rho_q = (1-2f_{\mathcal{Q}})3H^2 - \dfrac{f(\mathcal{Q})}{2} \label{rho_f(Q)}\\
    p_q = (1 + 2f_{\mathcal{Q}})(3H^2 + 2\dot{H}) + f(\mathcal{Q})^2 + 2\dot{f}_QH \label{p_f(Q)}
\end{eqnarray}
Now, we have defined the fractional density parameters as \eqref{Omegam} and \eqref{OmegaDE}. Following the same procedure mentioned in Sect.~\ref{Subsect:f(P)relations}, we have obtained the similar relations as described in Eqs. \eqref{modifiedfriedman1} - \eqref{p_mvalue}. The relation between the redshift and the scale factor $a(t)$ is defined in \eqref{redshift_a(t)}. The equation \eqref{a(t)value} describes how the universe's expansion rate follows a specific mathematical pattern known as a power law. This equation provides the value of the scale factor, which represents the relative expansion of the universe, as a function of time.

\subsection{\large Reconstruction of \texorpdfstring{$f(Q)$}{Lg} and dynamical analysis}\label{Sect:f(Q)_reconstruct}
\begin{figure}[htbp]
    \centering
    \includegraphics[width=0.6\textwidth]{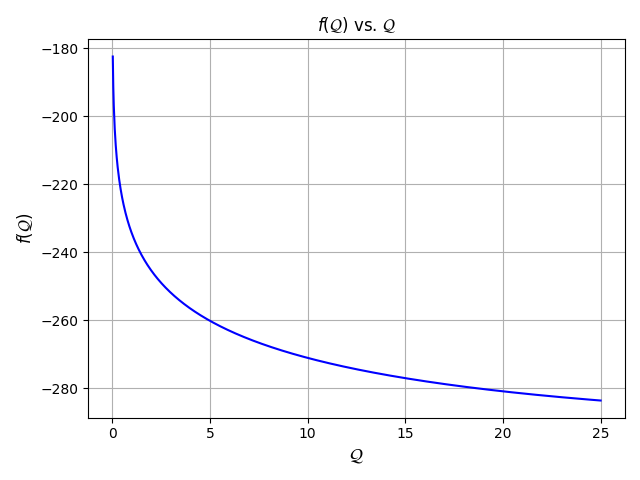}
    \caption{Plot of the function $f(\mathcal{Q})$ against the scalar $\mathcal{Q}$ for the reconstructed symmetric $f(\mathcal{Q})$ teleparallel gravity based on Barrow holographic dark energy.}
    \label{fig:f(Q)vsQ_plot}
\end{figure}

Here in this segment, we deduce the functional form of $f(\mathcal{Q})$ teleparallel gravity within the framework of Barrow holographic dark energy. The value of $Q$ can easily be obtained using Eqs. \eqref{Qvalue} and \eqref{H_delta} as,
\begin{eqnarray}\label{Q_delta}
    \mathcal{Q} = \dfrac{6H^2}{t^2}
\end{eqnarray}
The characteristic length given in Eq. \eqref{Lcalculatedvalue} is obtained in terms of $Q$ using Eqs. \eqref{a(t)value}, \eqref{H_delta}, \eqref{Q_delta} as,
\begin{equation}\label{L_from_f(Q)}
     L=\frac{M_2}{\mathcal{Q}^{2\epsilon}}
\end{equation}
Where, the values of $M_2$ and $\epsilon$ are given by $M_2=\frac{a_0^{(n-m)}(6\delta^2)^{2\epsilon}}{|1+\delta n|}$, $\epsilon=\frac{\delta(n-m)+1}{4}$. The term $n\delta + 1>0$ is associated with the expansion of the particle horizon, and $n\delta +1<0$ is associated with the contraction of the future event horizon. Then, we calculated the energy density of $f(\mathcal{Q})$ using Eqs. \eqref{rho_f(Q)}, \eqref{Q_delta}, \eqref{H_delta} as,
\begin{equation}\label{rho_Q_value}
    \rho_q = \frac{\mathcal{Q}}{2} - \mathcal{Q}f_{\mathcal{Q}} - \frac{f(\mathcal{Q})}{2}
\end{equation}
Now, to construct extended $f(\mathcal{Q})$ cubic gravity from Barrow holographic dark energy, we have to make an analogy by equating the dark energy densities given by,
\begin{equation}\label{f(Q)constructioneqtn}
  \frac{\mathcal{Q}}{2} - \mathcal{Q}f_{\mathcal{Q}} - \frac{f(\mathcal{Q})}{2}  = C \cdot \left( \frac{M_2}{\mathcal{Q}^{2 \epsilon}} \right)^{\Delta - 2}
\end{equation}
Where we have put $L=\frac{M_2}{Q^{2\epsilon}}$ in the equation~\eqref{rho_BHDE}. Solving this, we get, 
\begin{equation}\label{f(Q)value}
   \boxed{f(\mathcal{Q}) =  \frac{\mathcal{Q}}{3} + \frac{2C\mathcal{Q}^{4\epsilon}(M_2\mathcal{Q}^{-2\epsilon})^{\Delta}}{M_2^2(-1 + 4(-2 + \Delta)\epsilon)} + \frac{C_4}{\sqrt{\mathcal{Q}}}}
\end{equation}
Here, $C_4$ is the integration constant. The function $f(\mathcal{Q})$ is analytic except $\Delta = \dfrac{1+8\epsilon}{4\epsilon}$ and $M_2 = 0$ . This represents the required expression of the unknown function $f(\mathcal{Q})$ reconstructed from Barrow holographic dark energy. We have plotted the function \eqref{f(Q)value} against the non-metricity scalar $\mathcal{Q}$ in Fig.~\ref{fig:f(Q)vsQ_plot}. From the plot, we can conclude that $f(\mathcal{Q})$ decreases with $\mathcal{Q}$.

\section{{Methodology}}\label{Sect:Methodology}
We have employed Markov Chain Monte Carlo (MCMC) analysis to constraint the parameters of the reconstructed $f(\mathcal{P})$ and $f(\mathcal{Q})$ gravity models for the $(m,n)$-type Barrow holographic dark energy model. Our approach utilised a combined dataset comprising 26 Baryon Acoustic Oscillation (BAO) measurements and 31 Cosmic Chronometer (CC) data points, yielding a total of 57 observational constraints. Cosmic chronometers offer a straightforward approach to determine the Hubble parameter, denoted as $H(z)$, at different redshift values. This method involves estimating the age difference between passive galaxies, which are galaxies that have ceased active star formation and are composed primarily of older stellar populations. The key advantage of CC data is its independence from any specific cosmological model, making it a valuable tool for probing the expansion history of the universe. The CC data employed in this study consist of 31 measurements of $H(z)$ over a redshift range from $z \approx 0.07$ to $z \approx 1.97$. These measurements are compiled from various sources in the literature \cite{stern2010cosmic, moresco2012new, moresco2012improved, zhang2014four, moresco2015raising, moresco20166}, where the differential age method is applied to early-type galaxies. Baryon Acoustic Oscillations represent the imprint of sound waves in the early universe on the galaxy distributions. The BAO scale serves as a standard ruler, allowing for precise measurements of the angular diameter distance and the Hubble parameter at various redshifts. The BAO data utilised in this study consist of 26 measurements, including the isotropic and anisotropic BAO measurements from galaxy surveys such as SDSS \cite{ross2015clustering, alam2017clustering, gil2020completed, raichoor2021completed, hou2021completed, des2020completed}, Dark Energy Survey (DES) \cite{abbott2022dark}, Dark Energy Camera Legacy Survey (DECaLS) \cite{sridhar2020clustering}, 6dFGS BAO \cite{beutler20116df}. These measurements cover a broad redshift ranging from $0.106<z<2.36$. The complete datasets of CC+BAO is provided in table.~\ref{tab:cc_bao_dataset}.

\begin{table}[ht!]
\centering
\begin{minipage}{0.45\textwidth}
\centering
\begin{tabular}{c|c|c}
\toprule
$z$ & $H(z)$ [km/s/Mpc] & Error [km/s/Mpc] \\
\midrule
0.07   & 69.0  & 19.6  \\
0.09   & 69.0  & 12.0  \\
0.12   & 68.6  & 26.2  \\
0.17   & 83.0  & 8.0   \\
0.1791 & 75.0  & 4.0   \\
0.1993 & 75.0  & 5.0   \\
0.2    & 72.9  & 29.6  \\
0.24   & 79.69 & 2.99  \\
0.27   & 77.0  & 14.0  \\
0.28   & 88.8  & 36.6  \\
0.3    & 81.7  & 6.22  \\
0.31   & 78.18 & 4.74  \\
0.34   & 83.8  & 3.66  \\
0.35   & 84.7  & 9.1   \\
0.3519 & 83.0  & 14.0  \\
0.36   & 79.94 & 3.37  \\
0.38   & 81.5  & 1.9   \\
0.3802 & 83.0  & 13.5  \\
0.4    & 95.0  & 17.0  \\
0.4    & 82.04 & 2.03  \\
0.4004 & 77.0  & 10.2  \\
0.4247 & 87.1  & 11.2  \\
0.43   & 86.45 & 3.97  \\
0.44   & 82.6  & 7.8   \\
0.44   & 84.81 & 1.83  \\
0.4497 & 92.8  & 12.9  \\
\bottomrule
\end{tabular}
\end{minipage}%
\hfill
\begin{minipage}{0.45\textwidth}
\centering
\begin{tabular}{c|c|c}
\toprule
$z$ & $H(z)$ [km/s/Mpc] & Error [km/s/Mpc] \\
\midrule
0.47   & 89.0  & 34.0  \\
0.4783 & 80.9  & 9.0   \\
0.48   & 97.0  & 62.0  \\
0.48   & 87.79 & 2.03  \\
0.51   & 90.4  & 1.9   \\
0.52   & 94.35 & 2.64  \\
0.56   & 93.34 & 2.3   \\
0.57   & 87.6  & 7.8   \\
0.57   & 96.8  & 3.4   \\
0.59   & 98.48 & 3.18  \\
0.593  & 104.0 & 13.0  \\
0.6    & 87.9  & 6.1   \\
0.61   & 97.3  & 2.1   \\
0.64   & 98.82 & 2.98  \\
0.6797 & 92.0  & 8.0   \\
0.73   & 97.3  & 7.0   \\
0.7812 & 105.0 & 12.0  \\
0.8754 & 125.0 & 17.0  \\
0.88   & 90.0  & 40.0  \\
0.9    & 117.0 & 23.0  \\
1.037  & 154.0 & 20.0  \\
1.3    & 168.0 & 17.0  \\
1.363  & 160.0 & 33.6  \\
1.43   & 177.0 & 18.0  \\
1.53   & 140.0 & 14.0  \\
1.75   & 202.0 & 40.0  \\
1.965  & 186.5 & 50.4  \\
2.3    & 224.0 & 8.6   \\
2.33   & 224.0 & 8.0   \\
2.34   & 222.0 & 8.5   \\
2.36   & 226.0 & 9.3   \\
\bottomrule
\end{tabular}
\end{minipage}
\caption{Cosmic Chronometers and Baryon Acoustic Oscillation Dataset}
\label{tab:cc_bao_dataset}
\end{table}

\par Now, using eq. \eqref{friedmann1} one can derive the Hubble parameter relation as follows,
\begin{equation}\label{Hubble_Model_Relation}
    H(z) = H_0 \left[\Omega_{m0}(1+z)^{3(1+\omega_m)}+(1-\Omega_{m0})L^{\Delta-2}\right]^{\dfrac{1}{2}}
\end{equation}
Here, $\Omega_{m0} = \frac{\rho_{m0}}{3H_0^2}$ is the current value of matter density parameter, $\rho_{m0}$ is the density of matter in the current epoch, $\omega_m$ is the EoS parameter of matter, $L$ is the characteristic length of $(m,n)$-type BHDE and exact expression of $L$ is model specific as seen from the sect.~\ref{Sect:f(P)_reconstruct} and \ref{Sect:f(Q)_reconstruct}. The dark energy density parameter is written as $\Omega_{d0} = \frac{C}{3H_0^2}$ where, $C$ is a constant parameter defined in \eqref{rho_BHDE} and following the constraint relation on \eqref{modifiedfriedman1}, $\Omega_{d0}=1-\Omega_{m0}$. The expression in eq. \eqref{Hubble_Model_Relation} bridges the observational datasets of Hubble parameters with the theoretical frameworks of $f(\mathcal{P})$ and $f(\mathcal{Q})$ gravity models.
\par We implemented our MCMC algorithm using Python with \texttt{emcee} package \cite{foreman2013emcee}. The likelihood function was constructed assuming Gaussian errors in the observational data. Our prior distributions for the model parameters were chosen to be uniform priors within physically motivated bounds.
The MCMC chains were run for $8000$ steps after a burn-in period of $500$ steps to ensure convergence. We employed $100$ walkers to explore the parameter space efficiently. Convergence was assessed using the Gelman-Rubin statistic, ensuring that $R<1.1$ for all parameters. We have computed $\chi^2$ distance as, \[\chi^2 = \sum_{i=1}^{57} \frac{(O_i - T_i)^2}{\sigma_i^2}\] where $O_i$ denotes the observed value, $T_i$ represents the theoretical predictions and $\sigma_i$ represents the uncertainty associated with each data points. The joint likelihood function of the combined BAO + OHD(CC) datasets is defined as,  \[\mathcal{L}_{Total}=\mathcal{L}_{BAO}+\mathcal{L}_{CC}\] The outcomes of MCMC analysis is given in Tab~.\ref{tab:MCMCresults}.

\begin{table}[ht!]
\centering
\caption{MCMC Results and Priors for \( f(\mathcal{P}) \) and \( f(\mathcal{Q}) \) Gravity Models}
\renewcommand{\arraystretch}{1.5} % Adjust this value for more or less vertical spacing
\begin{tabular}{|p{3cm}|p{4cm}|p{5cm}|p{4cm}|}
\hline
\textbf{Model} & \textbf{Parameter} & \textbf{MCMC Result} & \textbf{Priors} \\ \hline

\multirow{5}{*}{\( f(\mathcal{P}) \)} 
 & \( H_{0} \) & \( 69.5240^{+0.68}_{-0.82} \)  & [60, 80] \\ \cline{2-4}
 & \( \Omega_{m0} \) & \( 0.2815^{+0.01}_{-0.02} \)  & [0.1, 0.3] \\ \cline{2-4}
 & \( \omega_m \) & \( -0.0091^{+0.02}_{-0.02} \)  & [-0.1, 1.5] \\ \cline{2-4}
 & \( \beta \) & \( -1.0935^{+0.77}_{-0.53} \)  & [-2.0, 2.0] \\ \cline{2-4}
 & \( \Delta \) & \( 0.4280^{+0.40}_{-0.27} \)  & [0, 1.0] \\ \hline
 
\multirow{4}{*}{\( f(\mathcal{Q}) \)} 
 & \( H_{0} \) & \( 69.6523^{+0.64}_{-0.76} \)  & [60, 80] \\ \cline{2-4}
 & \( \Omega_{m0} \) & \( 0.2752^{+0.02}_{-0.02} \)  & [0.1, 0.3] \\ \cline{2-4}
 & \( \omega_m \) & \( -0.0046^{+0.02}_{-0.02} \)  & [-0.1, 1.5] \\ \cline{2-4}
 & \( \Delta \) & \( 0.4464^{+0.39}_{-0.30} \)  & [0, 1.0] \\ \hline

\multirow{2}{*}{\( \Lambda CDM \)}
& \( H_{0} \) & \( 69.8195^{+1.06}_{-1.10} \)  & [50, 80] \\ \cline{2-4}
 & \( \Omega_{m0} \) & \( 0.2685^{+0.02}_{-0.02} \)  & [0.1, 0.4] \\ \hline
 
\end{tabular}
\label{tab:MCMCresults}
\end{table}
We analysed the posterior distributions of our model parameters to derive constraints on the (m,n)-type Barrow holographic dark energy. Credible intervals were computed using the 16th, 50th, and 84th percentiles of the marginalised posterior distributions. To visualise our results, we generated getdist plots showing the joint and marginalised posterior distributions for all parameters. We have fixed the values of parameters $C_2, C_3, C_4, n, m, a_0, \delta$ at $[0.0015, 10^7, 0.1, 3, 4, 0.68, 1.05]$ respectively. The reason for choosing $(m,n) =(4,3)$ lies in the fact that it produces good agreement with observation as mentioned in \cite{LING_2013}. We have used the best fit values given in table.~\ref{tab:MCMCresults} to obtain the plots described in the following section along with the above mentioned constant values.
\par When comparing a set of $(n)$ models, we employ specific criteria by calculating the relative difference of the involved information criterion (IC) values. This calculation is performed using the formula $(\Delta IC_{\text{model}} = IC_{\text{model}} - IC_{\text{min}})$, where $IC_{\text{min}}$ represents the minimum Information Criterion value among the models being compared \cite{anagnostopoulos2020observational}. According to the Jeffreys scale \cite{jeffreys1998theory}, the interpretation of the $\Delta IC$ values is as follows: a $\Delta IC \leq 2$ indicates that the model under consideration is statistically compatible with the reference model. A $\Delta IC$ value between 2 and 6 suggests a moderate tension between the two models, while a $\Delta IC \geq 10$ signifies a strong tension between them.

\begin{table}[ht!]
\centering
\begin{tabular}{|l|c|c|c|c|c|c|}
\hline
Model & AIC & \(\Delta \text{AIC}\) & BIC & \(\Delta \text{BIC}\) & DIC & \(\Delta \text{DIC}\) \\
\hline
\(\Lambda\text{CDM}\) & -28.62 & 0 & -24.54 & 0 & 36.15 & 0 \\
$f(\mathcal{P})$ & -21.43 & 7.19 & -16.72 & 7.82 & 36.42 & 0.27 \\
$f(\mathcal{Q})$ & -23.67 & 4.95 & -15.49 & 9.05 & 36.68 & 0.53 \\
\hline
\end{tabular}
\caption{AIC, BIC, and DIC values for the different models and their differences with respect to the \(\Lambda\text{CDM}\) model.}
\label{tab:model_metrics_comparison}
\end{table}

\begin{figure}[htbp]
    \centering
    \begin{subfigure}[b]{0.45\textwidth}
        \includegraphics[width=\textwidth]{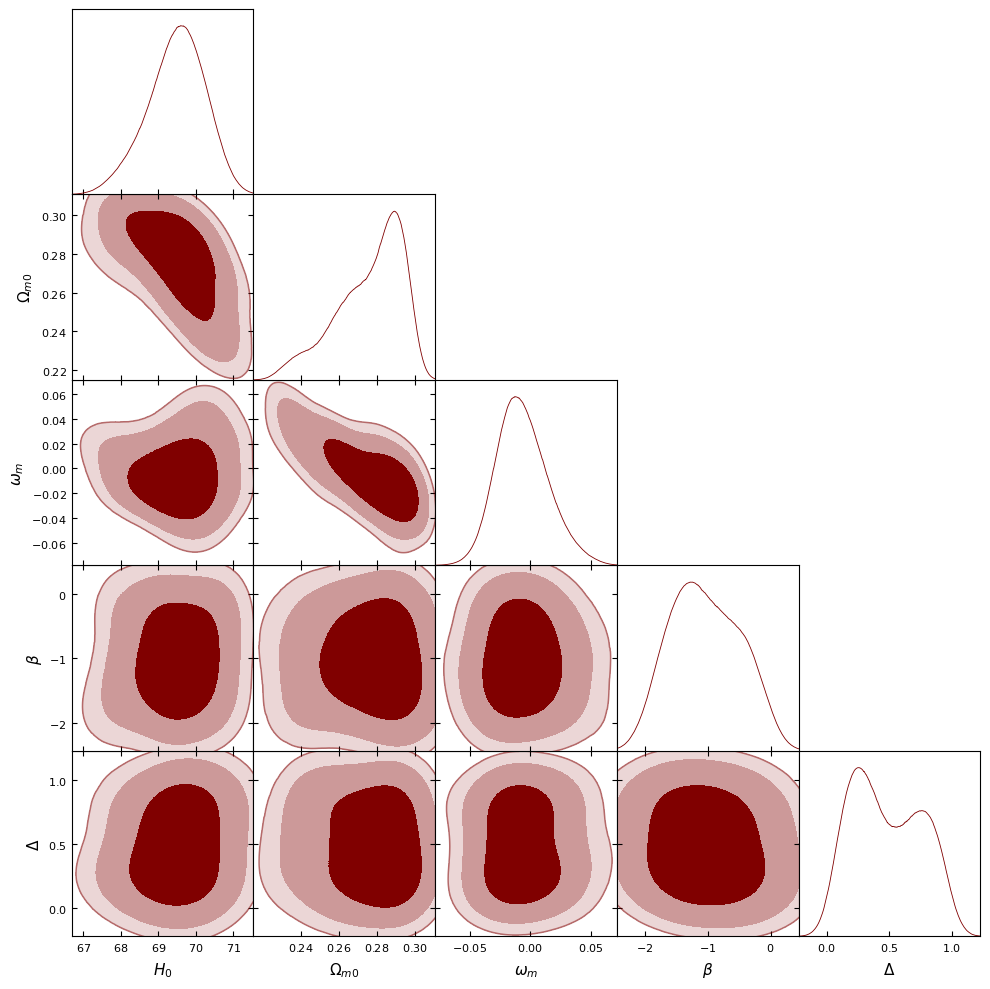}
        \caption{}
        \label{fig:getdist_f(P)}
    \end{subfigure}
    \begin{subfigure}[b]{0.45\textwidth}
        \includegraphics[width=\textwidth]{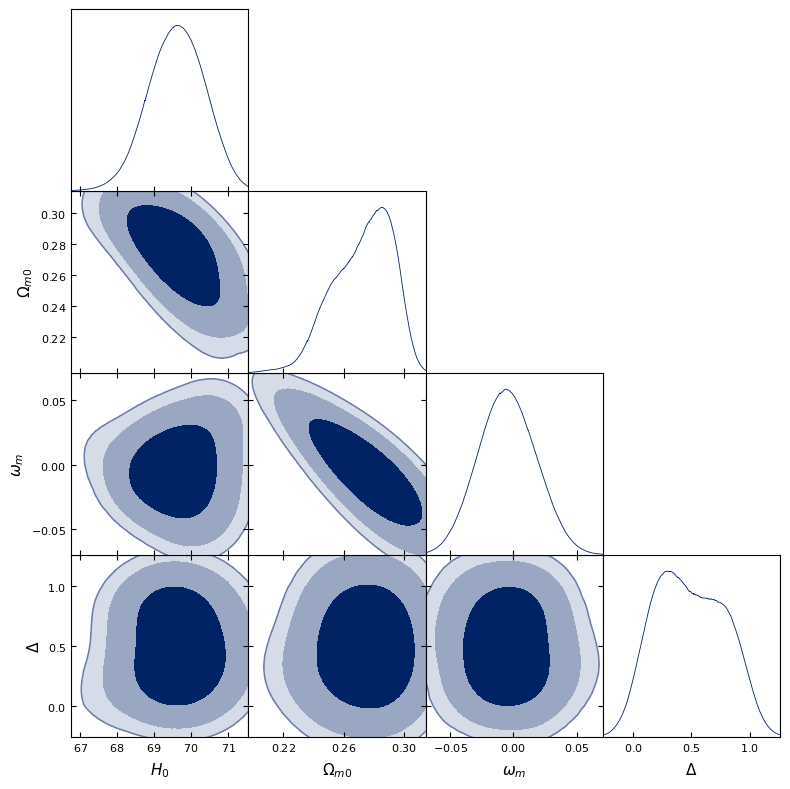}
        \caption{}
        \label{fig:getdist_f(Q)}
    \end{subfigure}
    \caption{Posterior distribution of $(m,n)$-type BHDE model parameters for fig: (a) $f(\mathcal{P})$ gravity and fig: (b) $f(\mathcal{Q})$ gravity at $1\sigma$ and $2\sigma$ confidence level.}
    \label{fig:getdist_f(P)_f(Q)}
\end{figure}

\section{{Results and Findings}}\label{Sect:Result}
\begin{figure}[htbp]
    \centering
    \begin{subfigure}[b]{0.45\textwidth}
        \includegraphics[width=\textwidth]{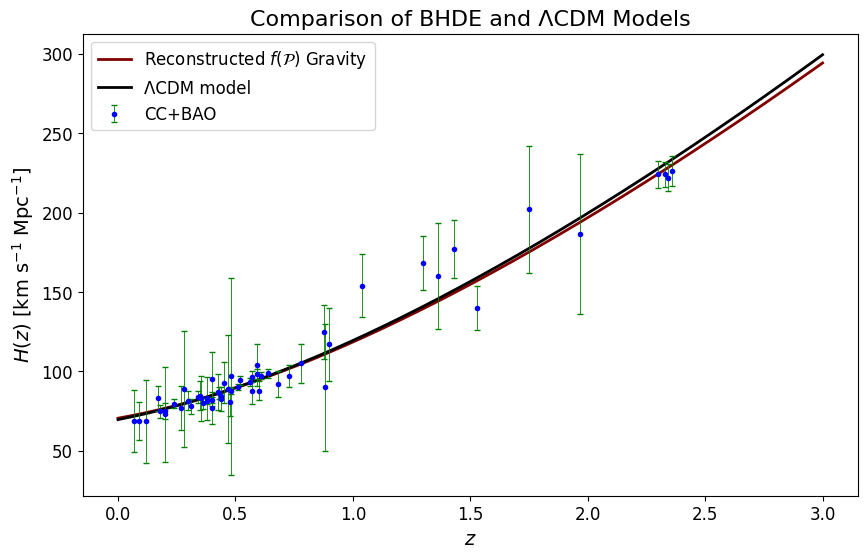}
        \caption{}
        \label{fig:Hubble_f(P)}
    \end{subfigure}
    \begin{subfigure}[b]{0.45\textwidth}
        \includegraphics[width=\textwidth]{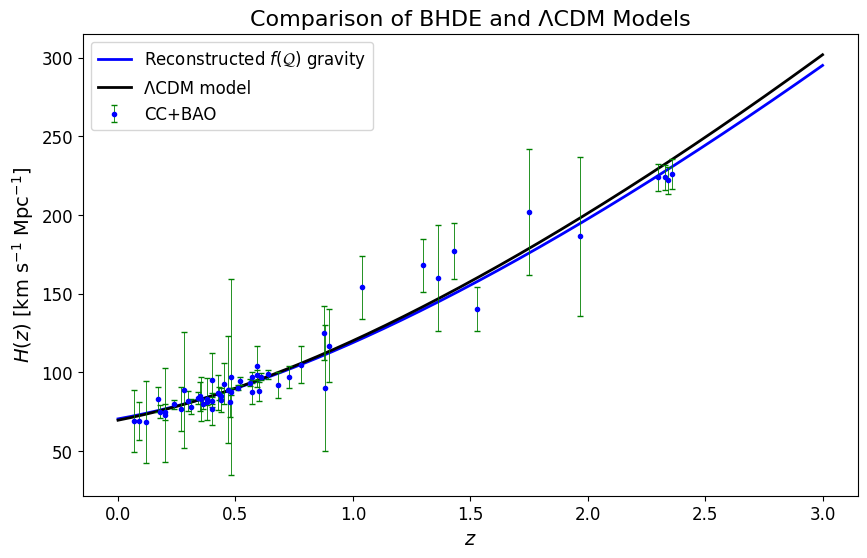}
        \caption{}
        \label{fig:Hubble_f(Q)}
    \end{subfigure}
    \caption{Comparison of Hubble Parameter between $\Lambda$CDM and the fig: (a) $f(\mathcal{P})$ gravity and fig: (b) $f(\mathcal{Q})$ gravity using 57 CC + BAO data points.}
    \label{fig:Hubble_f(P)_f(Q)}
\end{figure}

{\bf Comparison of Hubble Parameter Plot and Relative Difference:} The Hubble parameter values for the model of $f(\mathcal{P})$ and $f(\mathcal{Q})$ gravities are $69.5240^{+0.68}_{-0.82}$ km s$^{-1}$ Mpc$^{-1}$ and $69.6523^{+0.64}_{-0.76}$ km s$^{-1}$ Mpc$^{-1}$ respectively which is in alignment with the value obtained from Planck Collaboration \cite{refId0} but has a deviation with value obtained from SH0ES collaboration 2019 \cite{riess2019large}. In Fig.~\ref{fig:Hubble_f(P)_f(Q)}, we have presented the evolution of Hubble parameter $H(z)$ against the redshift $z$ for the $\Lambda$CDM model and the reconstructed $f(\mathcal{P})$, $f(\mathcal{Q})$ from $(m,n)$-type BHDE. The data points of the CC+BAO dataset are presented in blue. Both the plots in fig.~\ref{fig:Hubble_f(P)} and ~\ref{fig:Hubble_f(Q)} exhibit the well-fitted scenario of our models with the observed data. This agreement with the observed data establishes the capabilities of our models to explain the evolution dynamics of the universe. We have included the established $\Lambda$CDM model ($\Omega_{m0} = 0.3$ and $\Omega_{d0} = 0.7$) for the sake of better comparison. As evident from the fig.~\ref{fig:Hubble_f(P)_f(Q)}, the $f(\mathcal{P})$ and $f(\mathcal{Q})$ gravity models are in good agreement in lower redshift. As the redshift increases, the models start to deviate slightly from the $\Lambda$CDM paradigm.

\begin{figure}[htbp]
    \centering
    \begin{subfigure}[b]{0.45\textwidth}
        \includegraphics[width=\textwidth]{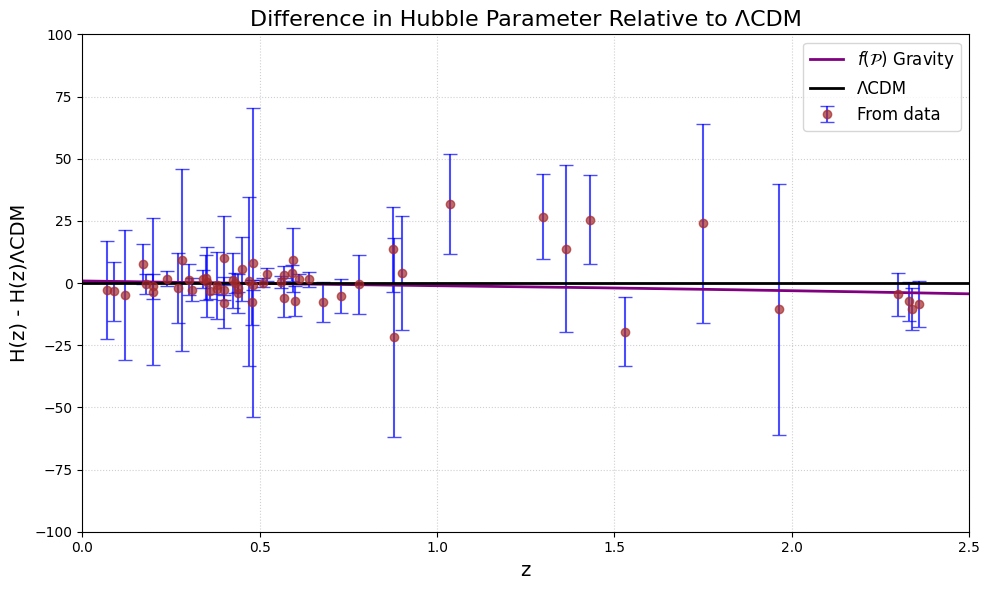}
        \caption{}
        \label{fig:Relative Hubble_f(P)}
    \end{subfigure}
    \begin{subfigure}[b]{0.45\textwidth}
        \includegraphics[width=\textwidth]{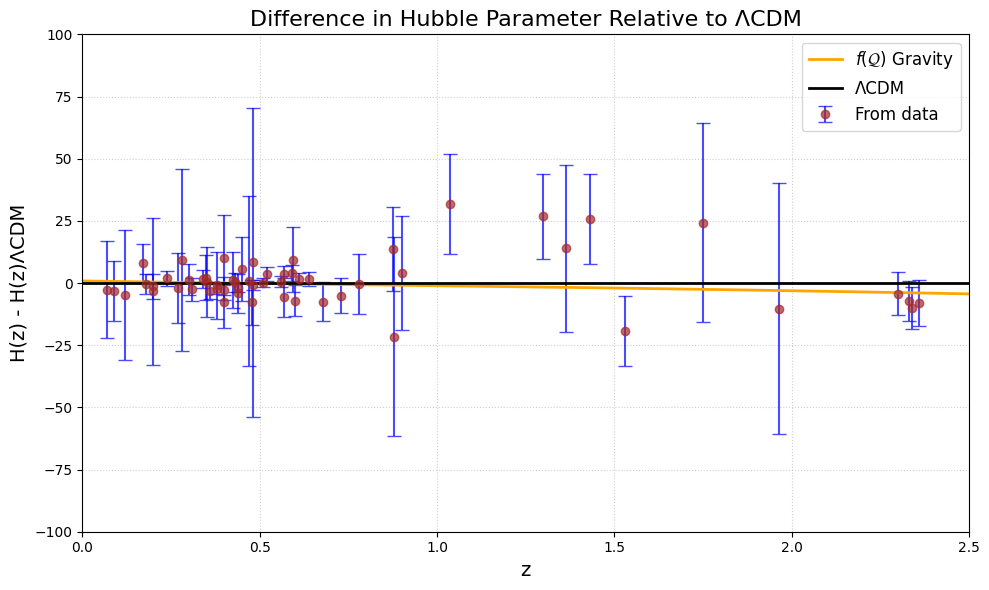}
        \caption{}
        \label{fig:Relative Hubble_f(Q)}
    \end{subfigure}
    \caption{Comparison of variation of difference in Hubble Parameter between $\Lambda$CDM and the fig: (a) $f(\mathcal{P})$ gravity and fig: (b) $f(\mathcal{Q})$ gravity using 57 CC + BAO data points.}
    \label{fig:Relative Hubble_f(P)_f(Q)}
\end{figure}

Fig.~\ref{fig:Relative Hubble_f(P)_f(Q)} illustrates the variation in the difference between the New Model in both frameworks and the ΛCDM model. For redshifts greater than 1 $( z > 1 )$, a distinct deviation of both frameworks becomes evident when compared to the cosmic chronometer (CC) and BAO measurements. This indicates that at higher redshifts, our models diverge slightly from the predictions made by the ΛCDM model, suggesting that the New Model may incorporate different physical processes or parameters that become more pronounced at these higher redshifts.
However, for redshifts less than 1 $( z <1 )$, this deviation diminishes. As the redshift decreases, the new models gradually become more consistent with the ΛCDM model. This suggests that at lower redshifts, the differences between the $f(\mathcal{P})$, $f(\mathcal{Q})$ gravity models and the ΛCDM model are less pronounced, and the predictions of both models align more closely with the observational data provided by the cosmic chronometers and Baryon Acoustic Oscillation.

\begin{figure}[htbp]
    \centering
    \begin{subfigure}[b]{0.45\textwidth}
        \includegraphics[width=\textwidth]{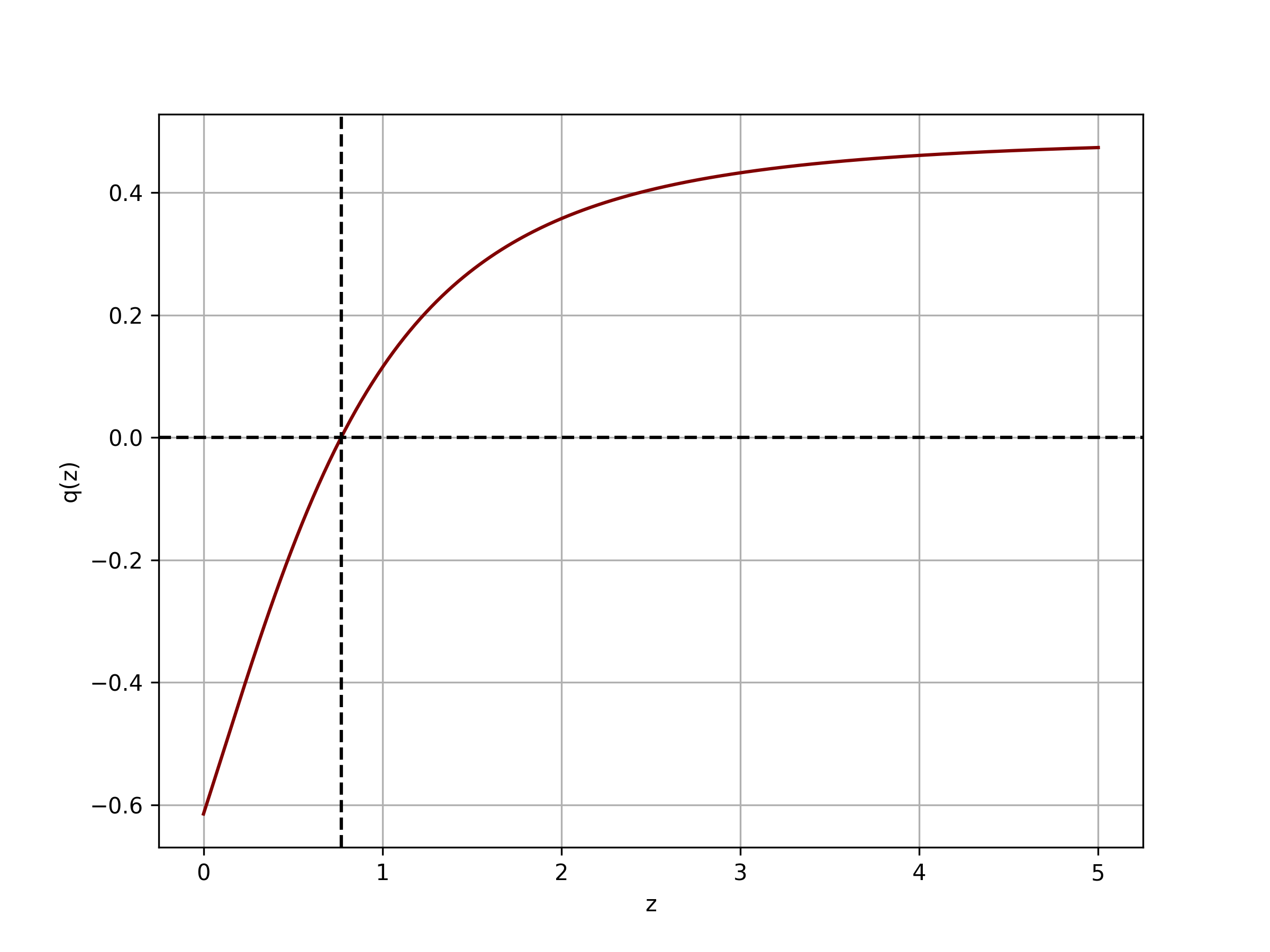}
        \caption{}
        \label{fig:q_f(P)}
    \end{subfigure}
    \begin{subfigure}[b]{0.45\textwidth}
        \includegraphics[width=\textwidth]{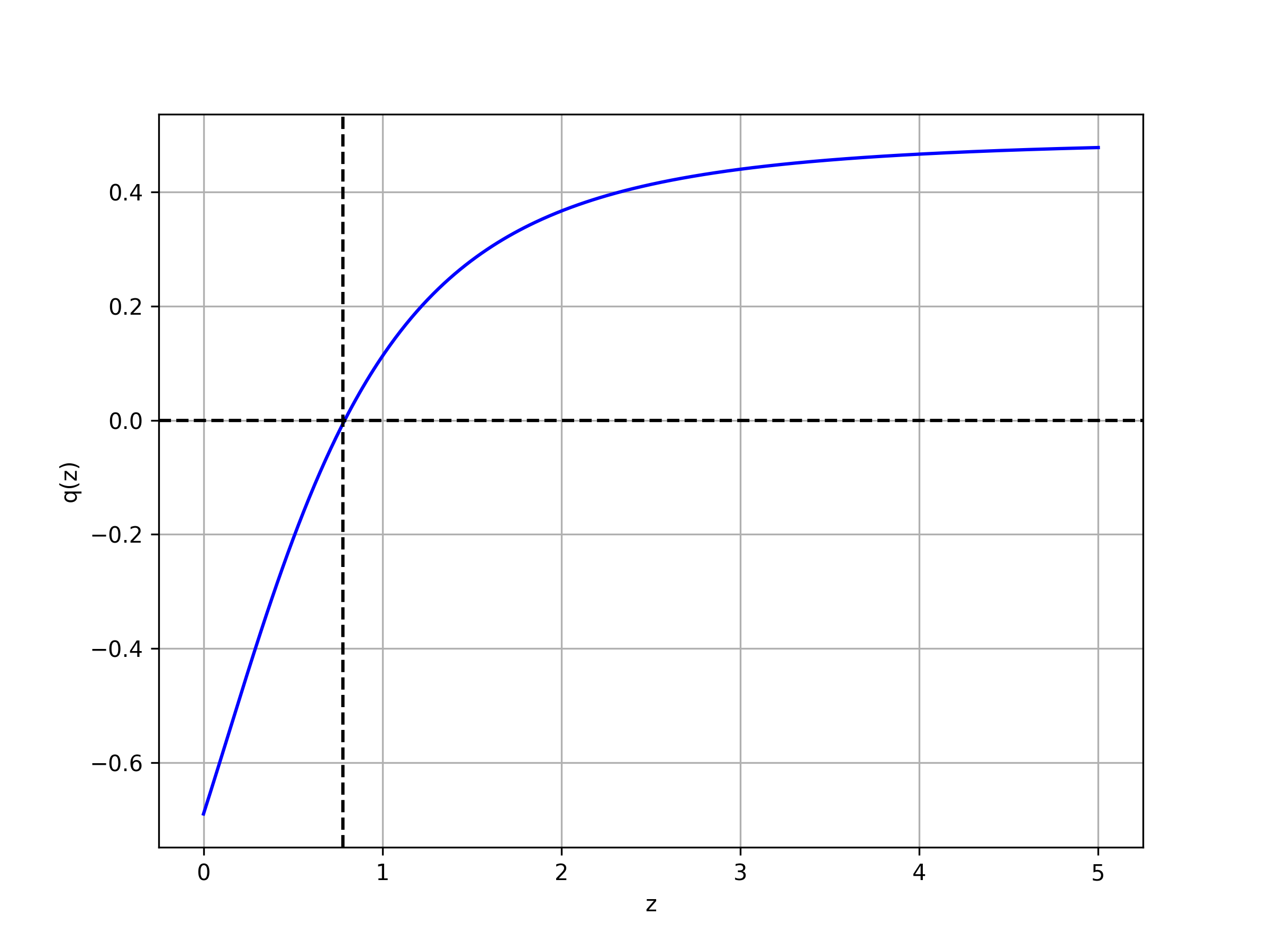}
        \caption{}
        \label{fig:q_f(Q)}
    \end{subfigure}
    \caption{Plot of evolution of deceleration parameter for the fig: (a) $f(\mathcal{P})$ gravity and fig: (b) $f(\mathcal{Q})$ gravity.}
    \label{fig:q_f(P)_f(Q)}
\end{figure}

\begin{figure}[htbp]
    \centering
    \begin{subfigure}[b]{0.45\textwidth}
        \includegraphics[width=\textwidth]{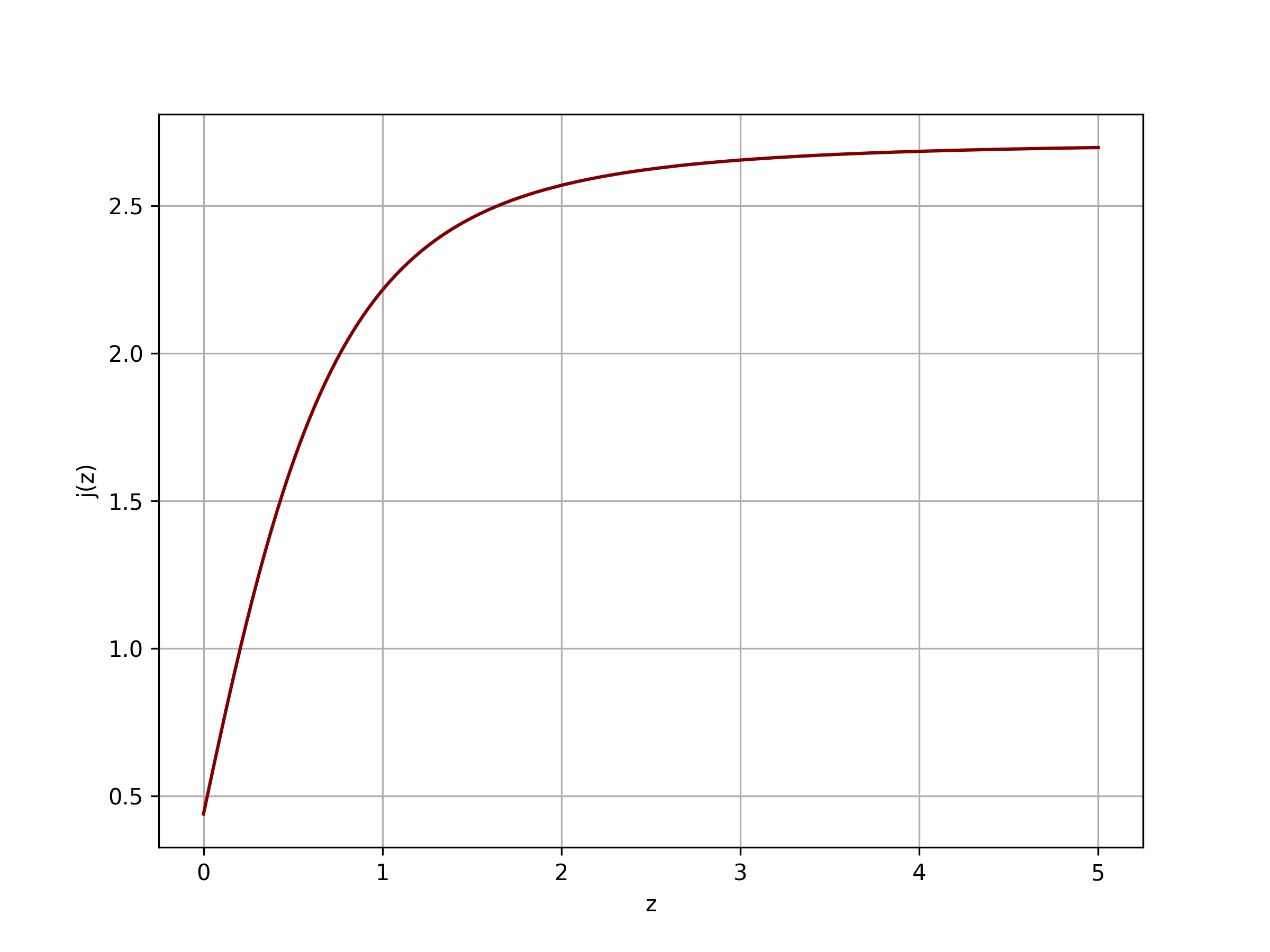}
        \caption{}
        \label{fig:j_f(P)}
    \end{subfigure}
    \begin{subfigure}[b]{0.45\textwidth}
        \includegraphics[width=\textwidth]{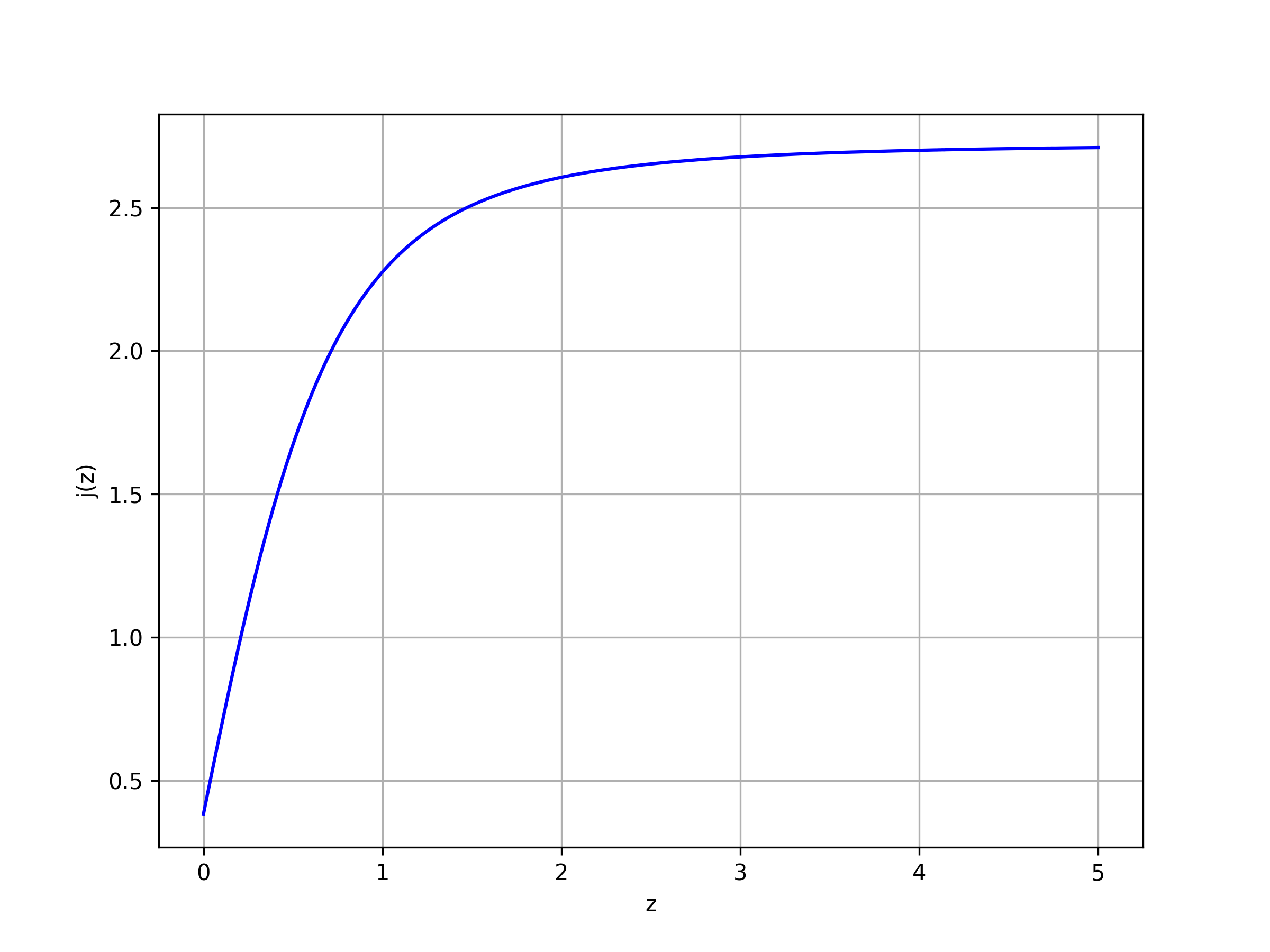}
        \caption{}
        \label{fig:j_f(Q)}
    \end{subfigure}
    \caption{Plot of evolution of jerk parameter for the fig: (a) $f(\mathcal{P})$ gravity and fig: (b) $f(\mathcal{Q})$ gravity.}
    \label{fig:j_f(P)_f(Q)}
\end{figure}

\begin{figure}[htbp]
    \centering
    \begin{subfigure}[b]{0.45\textwidth}
        \includegraphics[width=\textwidth]{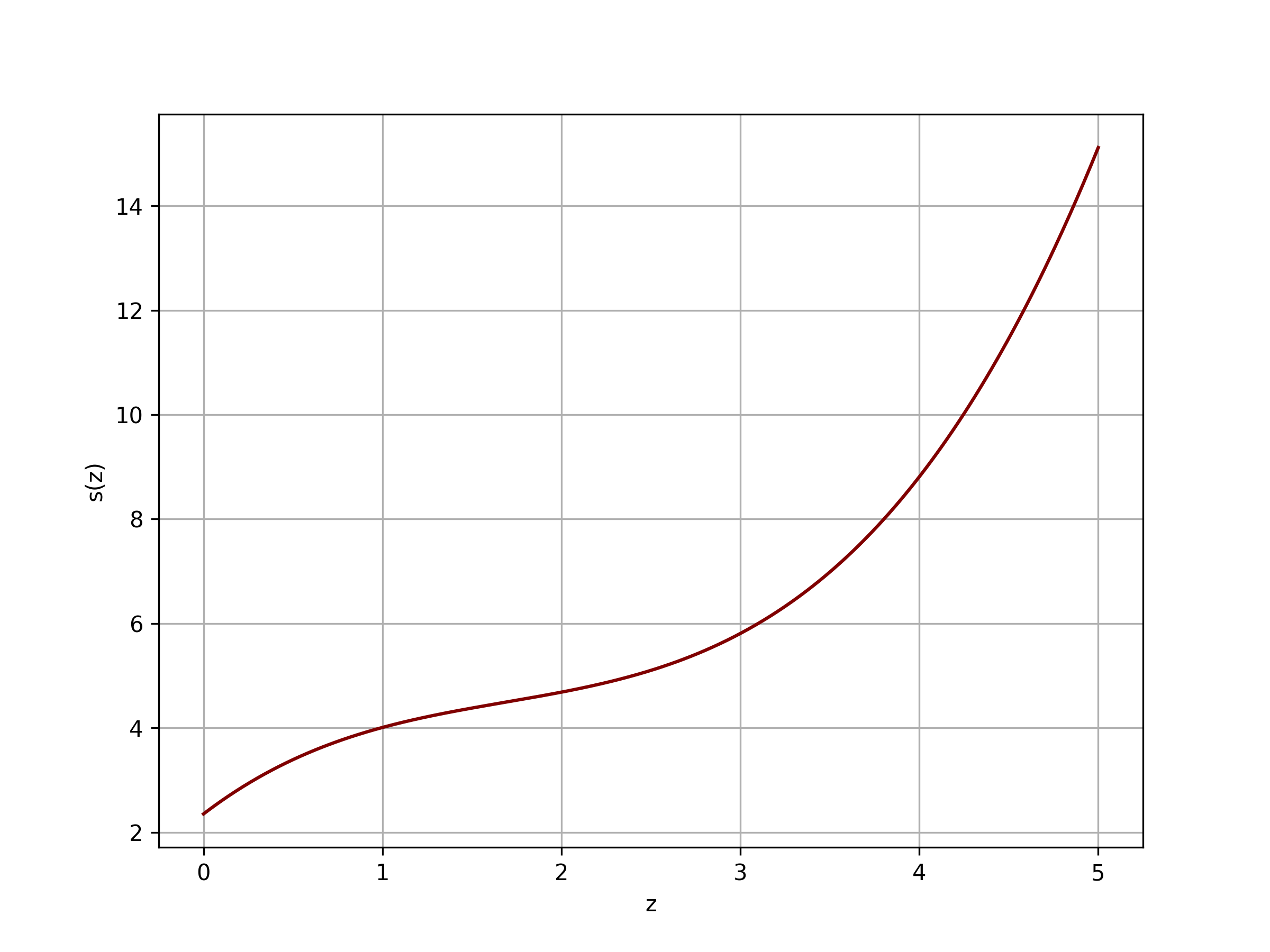}
        \caption{}
        \label{fig:s_f(P)}
    \end{subfigure}
    \begin{subfigure}[b]{0.45\textwidth}
        \includegraphics[width=\textwidth]{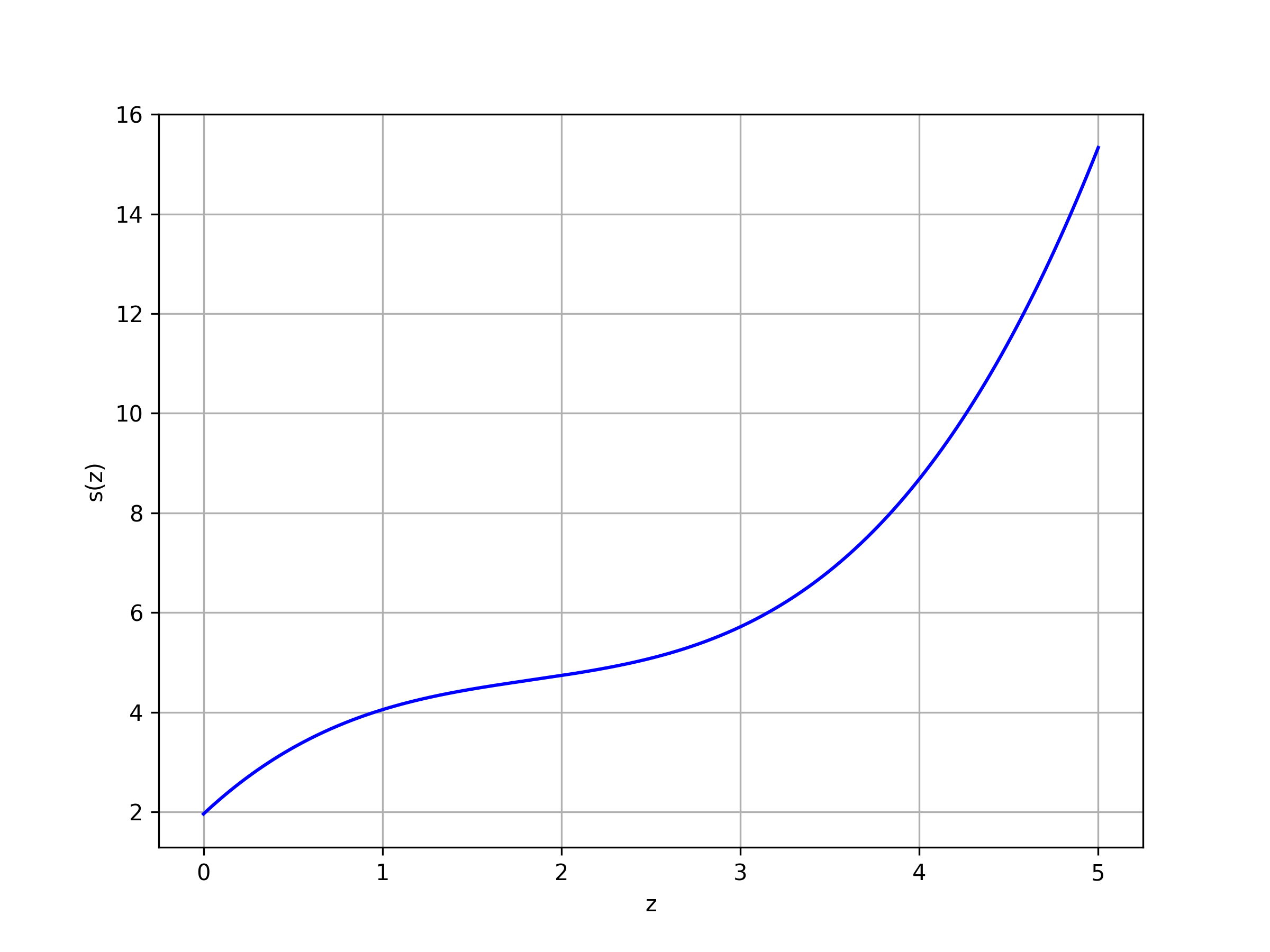}
        \caption{}
        \label{fig:s_f(Q)}
    \end{subfigure}
    \caption{Plot of evolution of snap parameter for the fig: (a) $f(\mathcal{P})$ gravity and fig: (b) $f(\mathcal{Q})$ gravity.}
    \label{fig:s_f(P)_f(Q)}
\end{figure}

\begin{figure}[htbp]
    \centering
    \begin{subfigure}[b]{0.45\textwidth}
        \includegraphics[width=\textwidth]{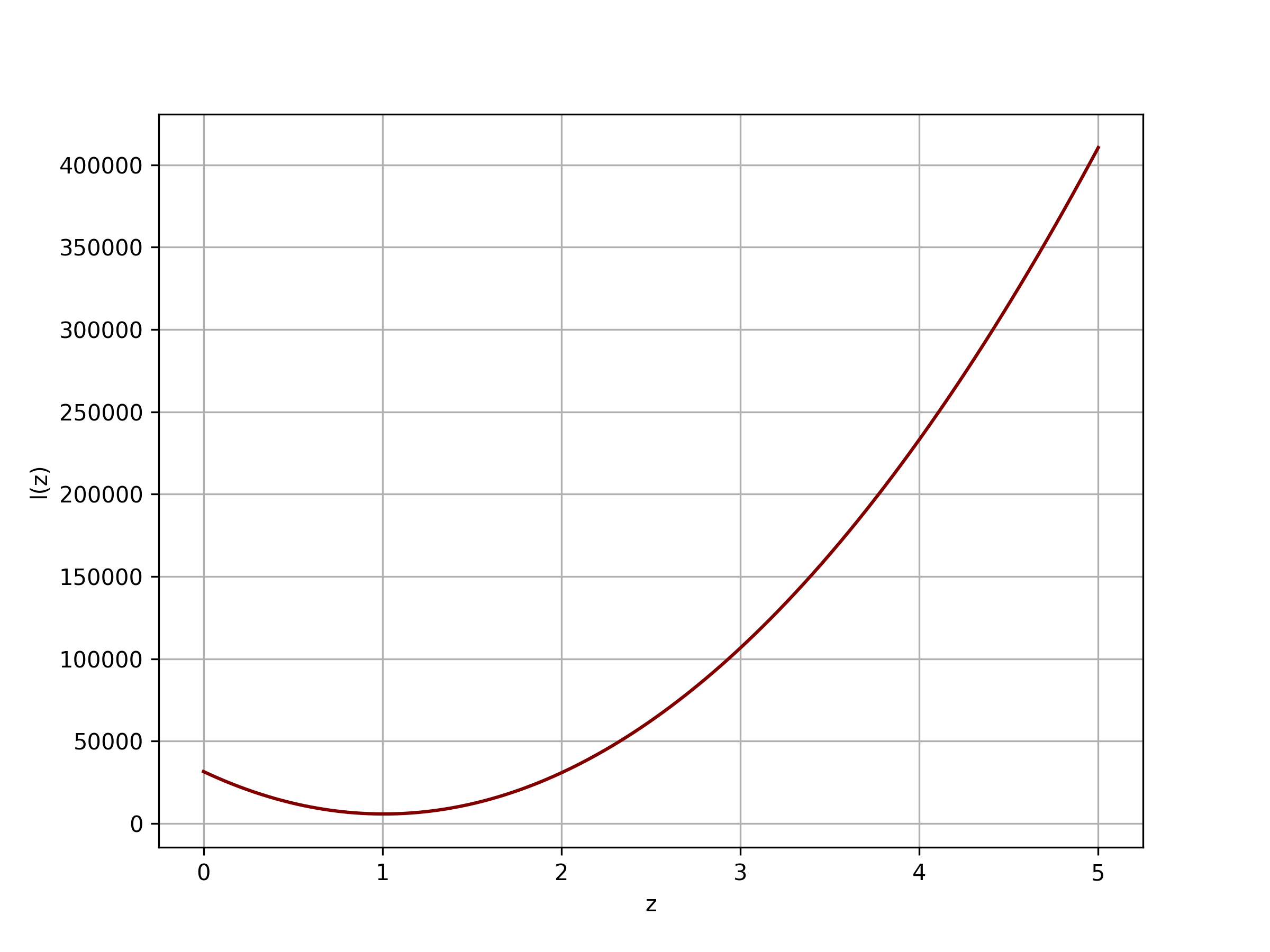}
        \caption{}
        \label{fig:l_f(P)}
    \end{subfigure}
    \begin{subfigure}[b]{0.45\textwidth}
        \includegraphics[width=\textwidth]{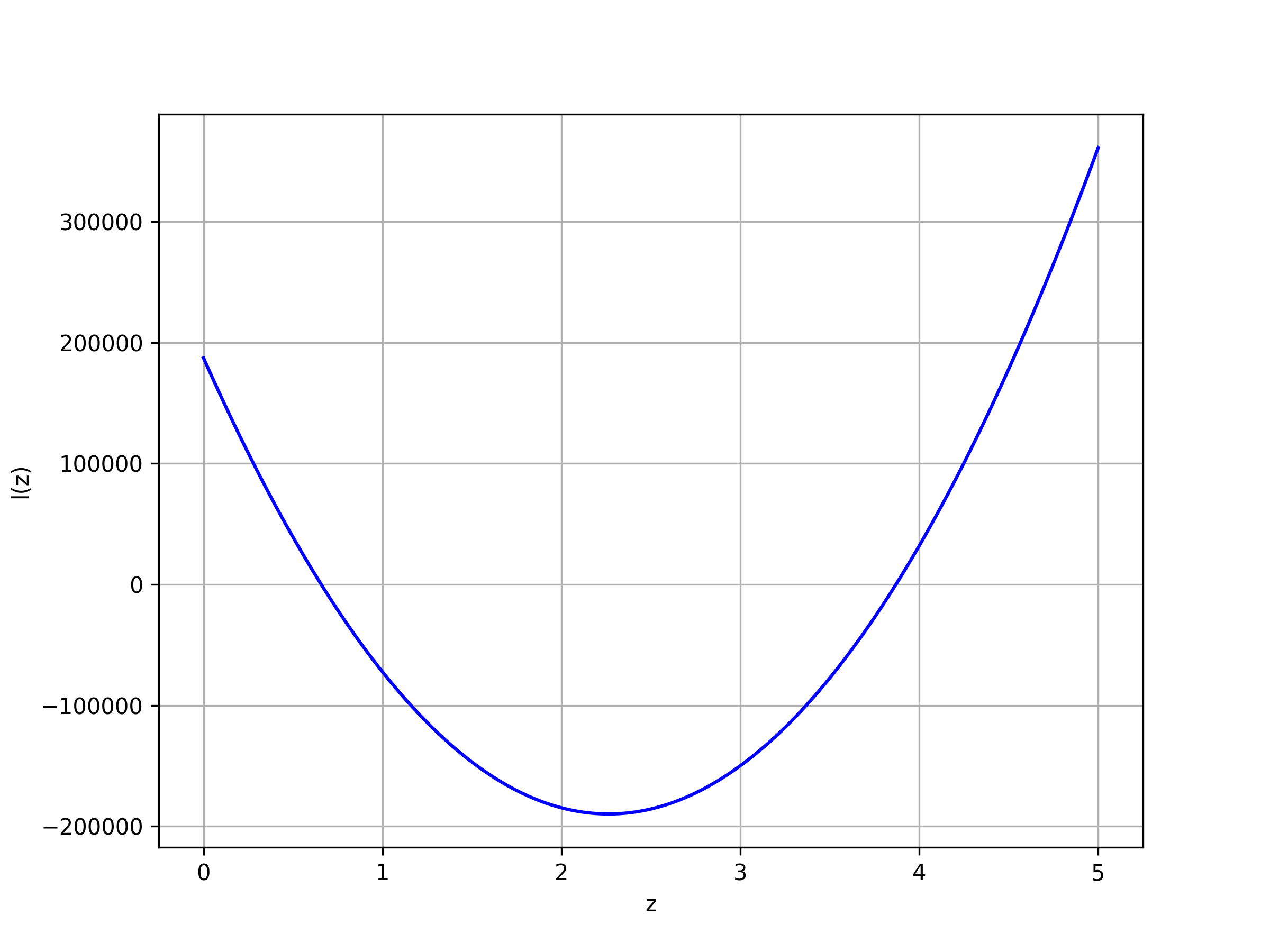}
        \caption{}
        \label{fig:l_f(Q)}
    \end{subfigure}
    \caption{Plot of evolution of lerk parameter for the fig: (a) $f(\mathcal{P})$ gravity and fig: (b) $f(\mathcal{Q})$ gravity.}
    \label{fig:l_f(P)_f(Q)}
\end{figure}

{\bf Cosmographic Parameters:} Fig.~\ref{fig:q_f(P)_f(Q)} illustrates the deceleration parameter’s $(q(z))$ trajectory, shifting from the decelerating phase to the accelerating phase around redshift $z_t \approx 0.7779$ for $f(\mathcal{P})$ and $z_t \approx 0.7891$ for $f(\mathcal{Q})$ gravity models, which are in close agreement with the observational values from SNIa + CMB + LSS joint analysis. The higher value of transition redshift $z_t$ in $f(\mathcal{Q})$ gravity indicates that the acceleration rate is different for both models. The current estimation of the deceleration parameter, denoted as $q(z=0)$, for the gravitational model based on the $f(\mathcal{P})$ function is approximately $q_0=-0.6215$. On the other hand, the corresponding value for the gravitational model derived from the $f(\mathcal{Q})$ function is around $q_0=-0.6903$.
In Fig.~\ref{fig:j_f(P)_f(Q)}, the jerk parameter shows positive values throughout the evolution for both models, which signifies the accelerated phase of the expansion rate of the universe is currently in a state of growth. The snap and lerk parameters in Fig.~\ref{fig:s_f(P)} and Fig.~\ref{fig:l_f(P)}  show positive values for the entire range of the redshift $z$ for $f(\mathcal{P})$ gravity whereas in Fig.~\ref{fig:s_f(Q)} and Fig.~\ref{fig:l_f(Q)} the snap parameter is positive but the lerk parameter becomes negative at certain redshift intervals and becomes positive again for $z<1$. It indicates the jerk parameter and the snap parameter are increasing over the evolution. This pattern suggests that, over time, the rate of acceleration of the universe’s expansion is on the rise and the rate of change of acceleration is also increasing.

\begin{figure}[htbp]
    \centering
    \begin{subfigure}[b]{0.45\textwidth}
        \includegraphics[width=\textwidth]{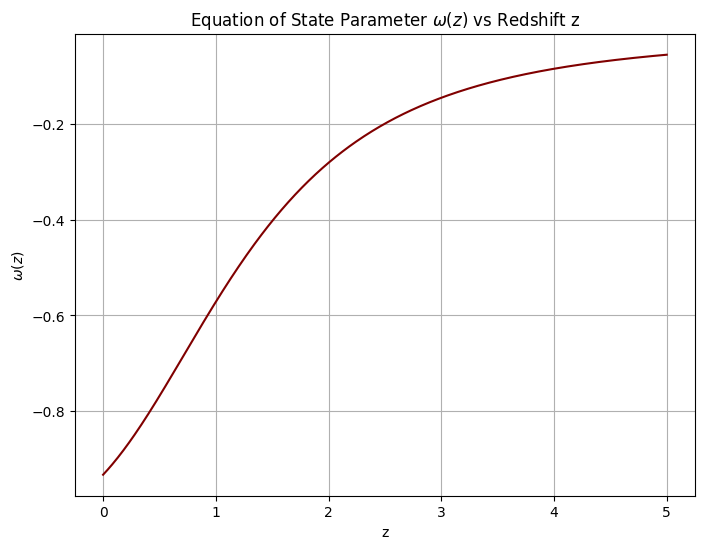}
        \caption{}
        \label{fig:omega_f(P)}
    \end{subfigure}
    \begin{subfigure}[b]{0.45\textwidth}
        \includegraphics[width=\textwidth]{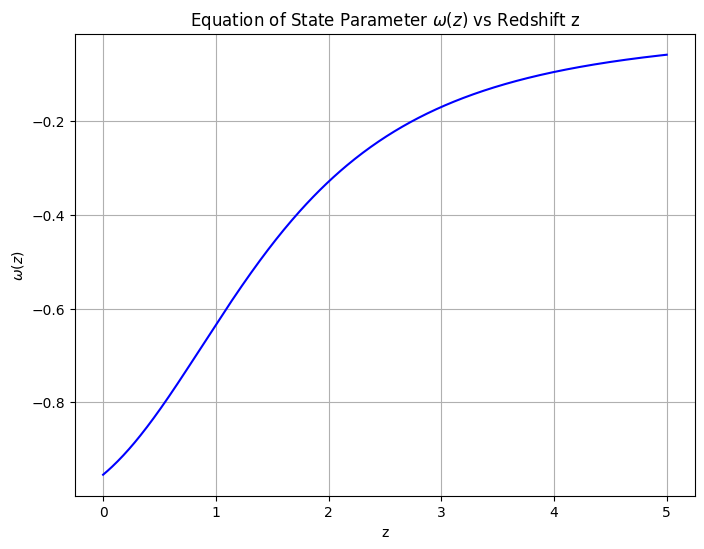}
        \caption{}
        \label{fig:omega_f(Q)}
    \end{subfigure}
    \caption{Plot of evolution trajectory of EoS parameter for dark energy $\omega_{DE}(z)$ for the fig: (a) $f(\mathcal{P})$ gravity and fig: (b) $f(\mathcal{Q})$ gravity.}
    \label{fig:omega_f(P)_f(Q)}
\end{figure}

\begin{figure}[htbp]
    \centering
    \begin{subfigure}[b]{0.45\textwidth}
        \includegraphics[width=\textwidth]{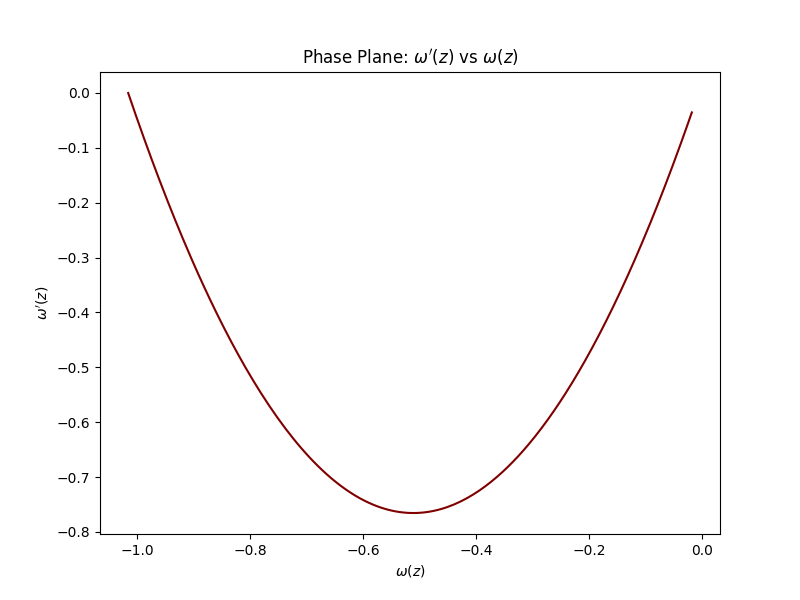}
        \caption{}
        \label{fig:omega'_f(P)}
    \end{subfigure}
    \begin{subfigure}[b]{0.45\textwidth}
        \includegraphics[width=\textwidth]{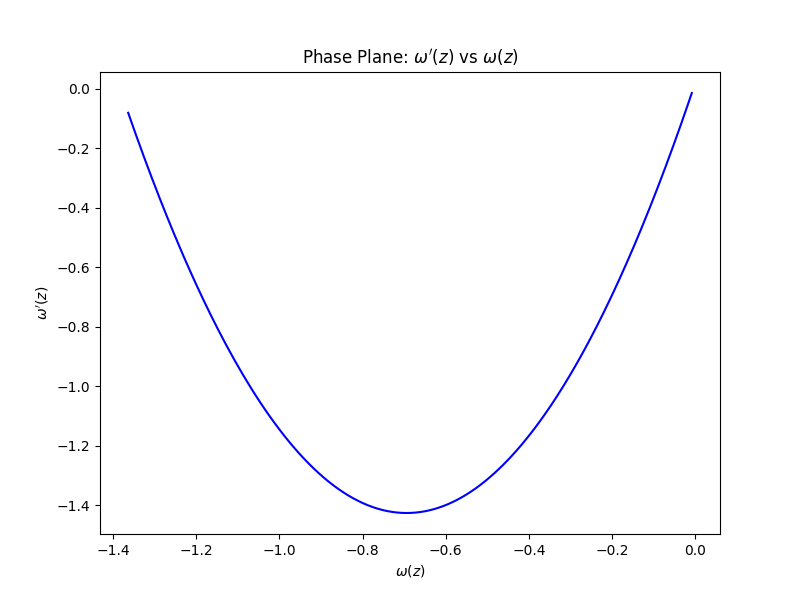}
        \caption{}
        \label{fig:omega'_f(Q)}
    \end{subfigure}
    \caption{Plot of $\omega'_{DE}-\omega_{DE}$ phase plane for the fig: (a) $f(\mathcal{P})$ gravity and fig: (b) $f(\mathcal{Q})$ gravity.}
    \label{fig:omega'_f(P)_f(Q)}
\end{figure}

{\bf EoS Parameter and the $\omega'_{DE}-\omega_{DE}$ Phase Plane:} Fig.~\ref{fig:omega_f(P)_f(Q)} shows the evolution trajectory of the dark energy EoS parameter $\omega_{DE}$ which is negative throughout the evolution of the universe but increases with the decrease in redshift $z$, which indicates the beginning of dark energy dominated era. The plot suggests the presence of quintessence type dark energy for both models. The value of $\omega_{DE}$ at the current epoch is $\omega_{DE0}=-0.9328$ for $f(\mathcal{P})$ gravity and $\omega_{DE0}=-0.9542$ for $f(\mathcal{Q})$ gravity which is very close to the current observed value \cite{alam2021completed}.
The phase plane of $\omega'_{DE}-\omega_{DE}$ in Fig.~\ref{fig:omega'_f(P)_f(Q)} shows a freezing region for both models, signifying an increased acceleration in the universe’s expansion. 

\begin{figure}[htbp]
    \centering
    \begin{subfigure}[b]{0.45\textwidth}
        \includegraphics[width=\textwidth]{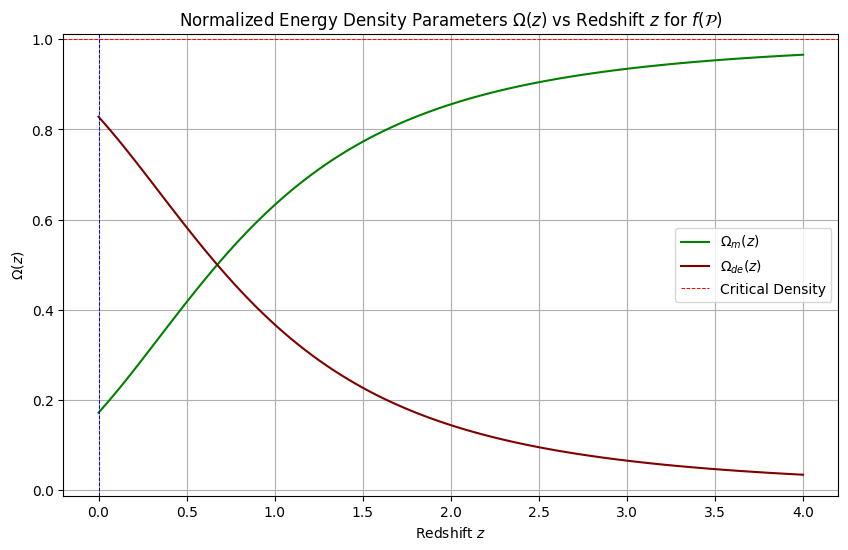}
        \caption{}
        \label{fig:Omega_total_f(P)}
    \end{subfigure}
    \begin{subfigure}[b]{0.45\textwidth}
        \includegraphics[width=\textwidth]{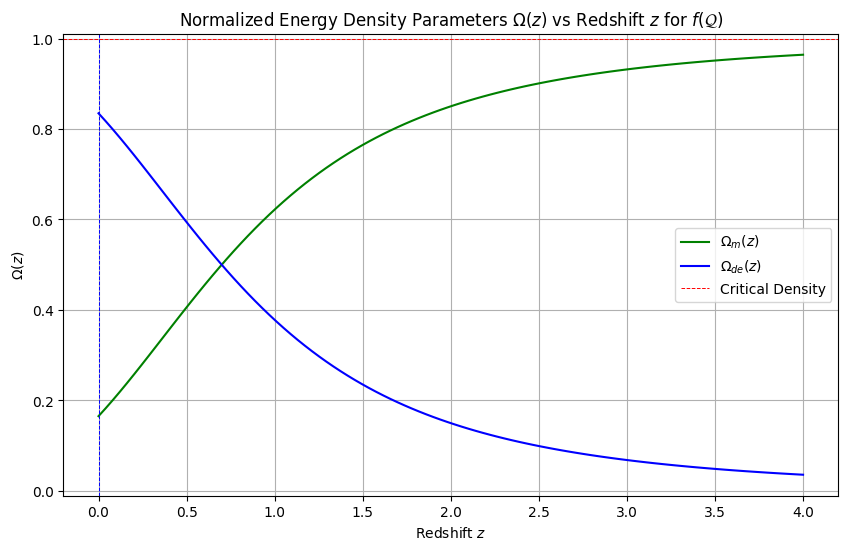}
        \caption{}
        \label{fig:Omega_total_f(Q)}
    \end{subfigure}
    \caption{Plot of density parameter components of matter and dark energy for the fig: (a) $f(\mathcal{P})$ gravity and fig: (b) $f(\mathcal{Q})$ gravity.}
    \label{fig:Omega_total_f(P)_f(Q)}
\end{figure}

{\bf Density Parameter:} The fig~.\ref{fig:Omega_total_f(P)_f(Q)} presents the evolution of normalised energy density parameters Ω(z) as a function of redshift z for our cosmological model of $f(\mathcal{P})$ and $f(\mathcal{Q})$ gravity. Two primary components are depicted: the matter density parameter $\Omega_m(z)$ and the dark energy density parameter $\Omega_{de}(z)$. The critical density is represented by a vertical red dashed line at z=0. The matter density parameter increases as we go back in time (higher redshift), indicating that the universe was more matter-dominated in the past. The dark energy density parameter decreases with increasing redshift, showing that dark energy becomes less dominant in the past and more dominant in today's universe. The curves cross at around $z \approx 0.7$, indicating a transition from a matter-dominated to a dark energy-dominated universe. The sum of $\Omega_m$ and $\Omega_{de}$ appears to be 1 at all times, consistent with a flat universe model. 

\begin{figure}[htbp]
    \centering
    \begin{subfigure}[b]{0.45\textwidth}
        \includegraphics[width=\textwidth]{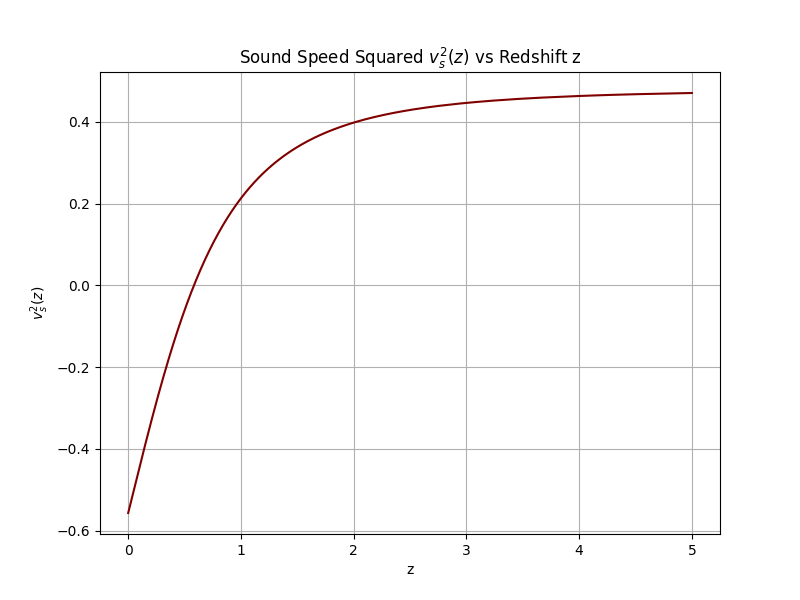}
        \caption{}
        \label{fig:v^2_f(P)}
    \end{subfigure}
    \begin{subfigure}[b]{0.45\textwidth}
        \includegraphics[width=\textwidth]{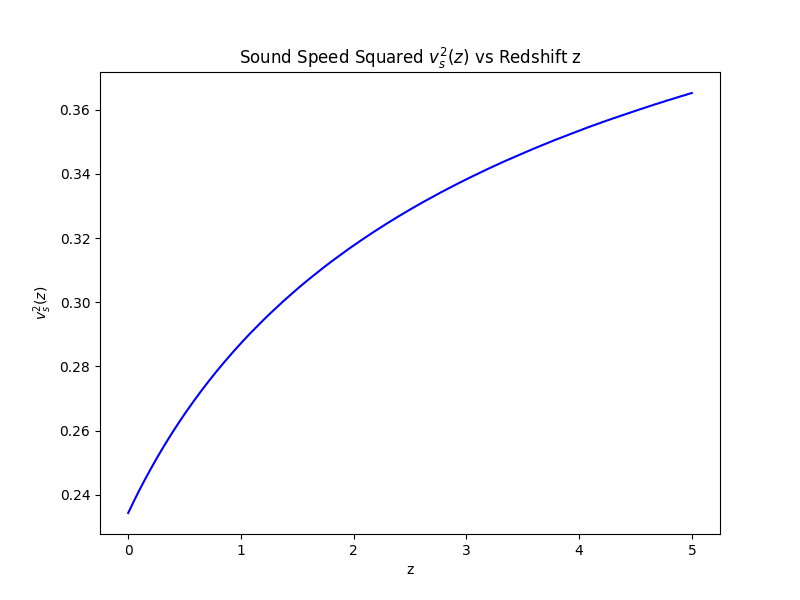}
        \caption{}
        \label{fig:v^2_f(Q)}
    \end{subfigure}
    \caption{Plot of square of the speed of sound $v_s^2$ for the fig: (a) $f(\mathcal{P})$ gravity and fig: (b) $f(\mathcal{Q})$ gravity.}
    \label{fig:v^2_f(P)_f(Q)}
\end{figure}

{\bf Square of the speed of sound:} The plot in Fig.~\ref{fig:v^2_f(P)_f(Q)} illustrates that the square of the velocity of sound $v_s^2$ is positive for $z>1$ and negative for $z<1$ for $f(\mathcal{P})$ gravity and for the $f(\mathcal{Q})$ gravity it is positive for the entire redshift interval. Thus, the reconstructed $f(\mathcal{Q})$ from $(m,n)$-type BHDE is classically stable under perturbation throughout the evolution of the universe, and the reconstructed $f(\mathcal{P})$ gravity is stable for higher redshift but unstable for lower redshift values.

\begin{figure}[htbp]
    \centering
    \begin{subfigure}[b]{0.45\textwidth}
        \includegraphics[width=\textwidth]{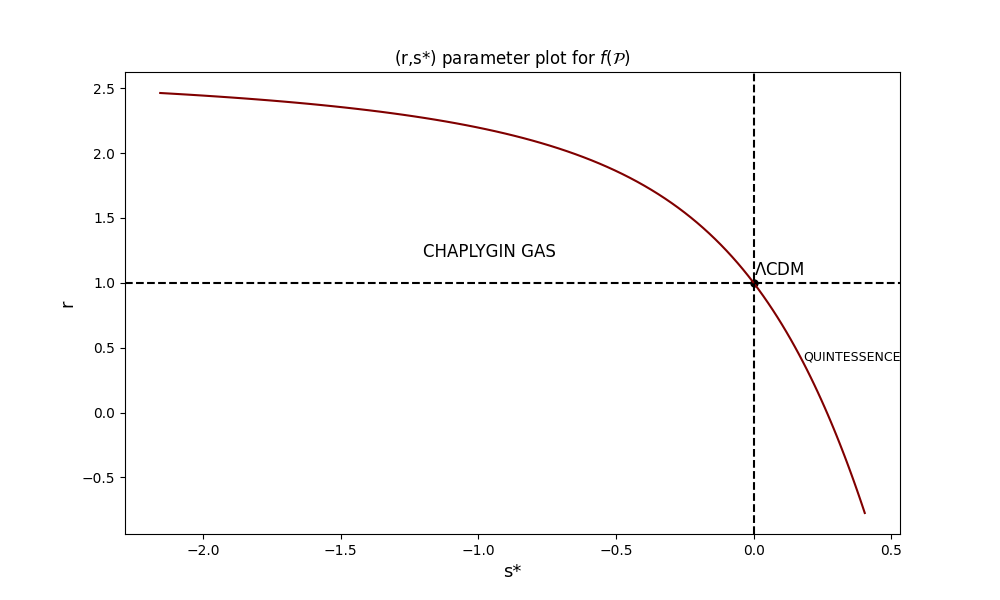}
        \caption{}
        \label{fig:r_s_f(P)}
    \end{subfigure}
    \begin{subfigure}[b]{0.45\textwidth}
        \includegraphics[width=\textwidth]{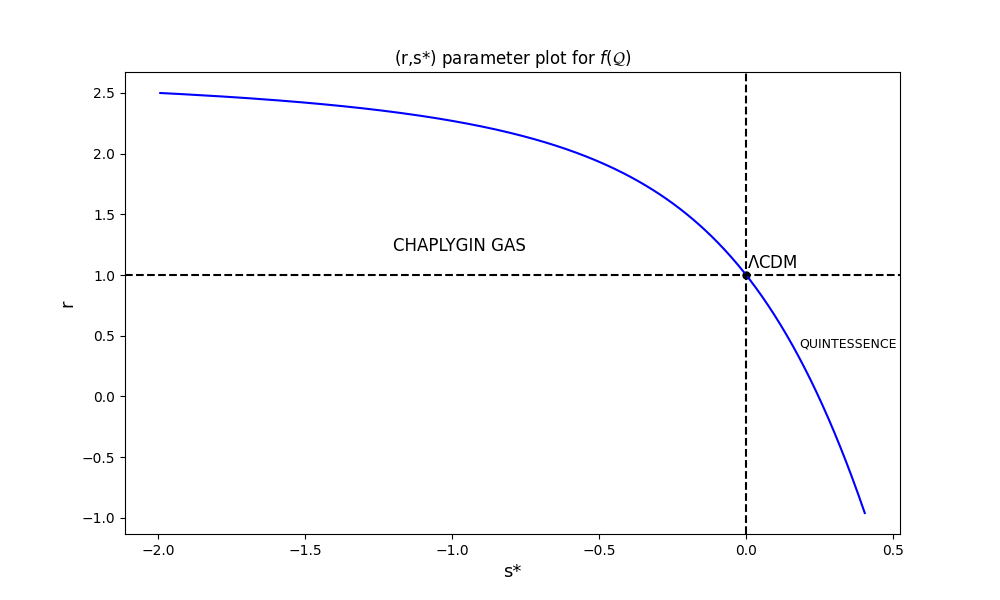}
        \caption{}
        \label{fig:r_s_f(Q)}
    \end{subfigure}
    \caption{Plot of evolution trajectory of $(r,s^*)$ statefinder parameter for the fig: (a) $f(\mathcal{P})$ gravity and fig: (b) $f(\mathcal{Q})$ gravity.}
    \label{fig:r_s_f(P)_f(Q)}
\end{figure}

\begin{figure}[htbp]
    \centering
    \begin{subfigure}[b]{0.45\textwidth}
        \includegraphics[width=\textwidth]{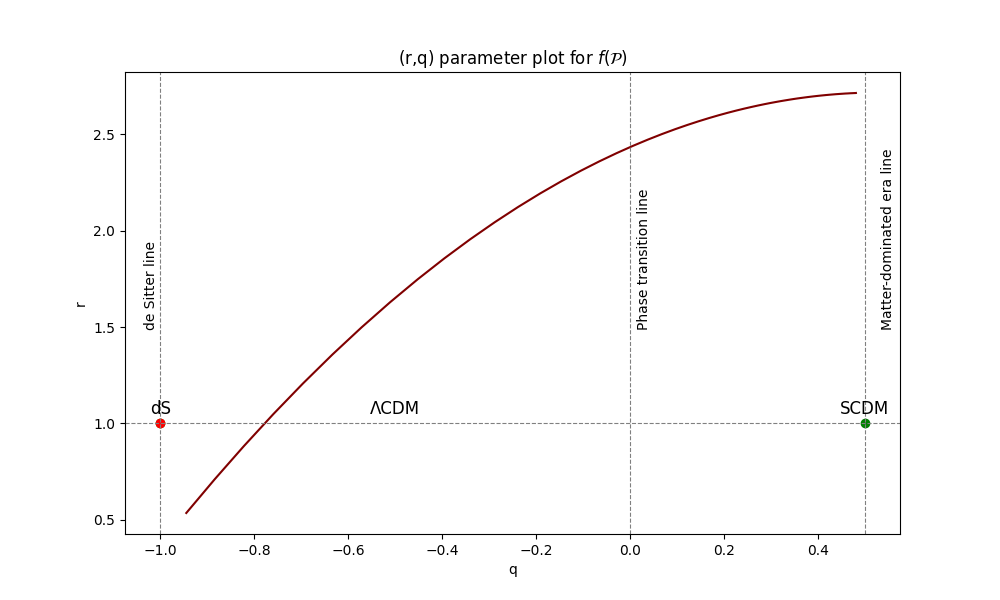}
        \caption{}
        \label{fig:r_q_f(P)}
    \end{subfigure}
    \begin{subfigure}[b]{0.45\textwidth}
        \includegraphics[width=\textwidth]{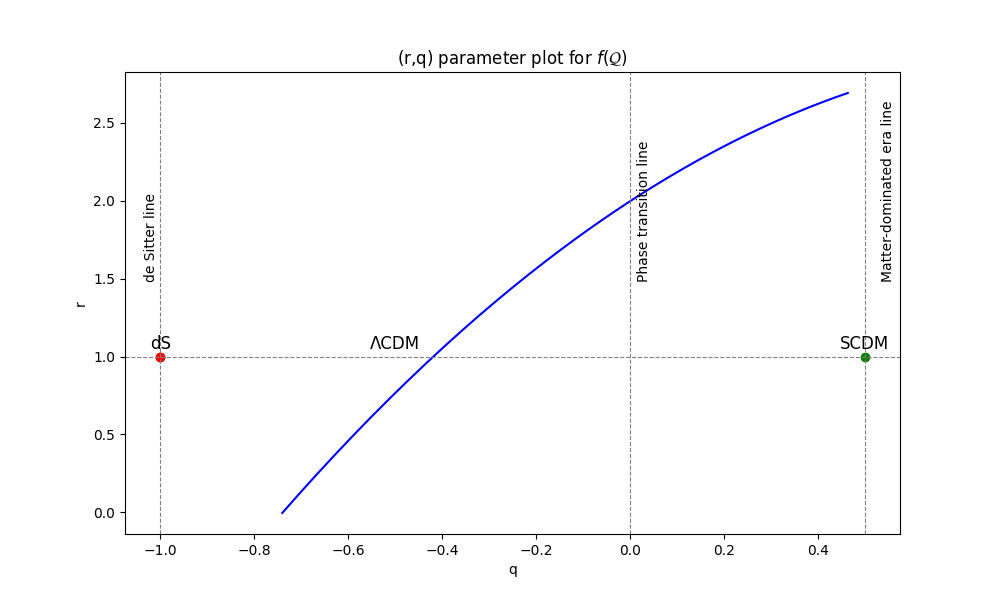}
        \caption{}
        \label{fig:r_q_f(Q)}
    \end{subfigure}
    \caption{Plot of evolution trajectory of $(r,q)$ statefinder parameter for the fig: (a) $f(\mathcal{P})$ gravity and fig: (b) $f(\mathcal{Q})$ gravity.}
    \label{fig:r_q_f(P)_f(Q)}
\end{figure}

{\bf Statefinder parameter:} In Fig.~\ref{fig:r_s_f(P)_f(Q)} the $(r,s^*)$ parameter plot shows that in the early stages within the framework, both model show values where $( r > 1 )$ and $( s^* < 0 )$, indicating it is in the quintessence region. As the models evolve, they transit and cross the fixed point at $(r, s^*) = (1, 0)$, which represents the ΛCDM model. After this transition, the model adopts values of $( r < 1 )$ and $( s^* > 0 )$, corresponding to the Chaplygin gas region. The $(r,q)$ plot in Fig.~\ref{fig:r_q_f(P)_f(Q)} illustrates the acceleration in the universe’s expansion, signifying a transition from an era dominated by matter to one governed by dark energy, though it does not attain the steady state nature $(r =1, q= -1)$ and SCDM $(r = 1, q = \frac{1}{2})$ for the reconstructed $f(\mathcal{P})$ and $f(\mathcal{Q})$ model. Both the models converge to the de Sitter line $q = -1$ in the far future.

\begin{figure}[htbp]
    \centering
    \begin{subfigure}[b]{0.45\textwidth}
        \includegraphics[width=\textwidth]{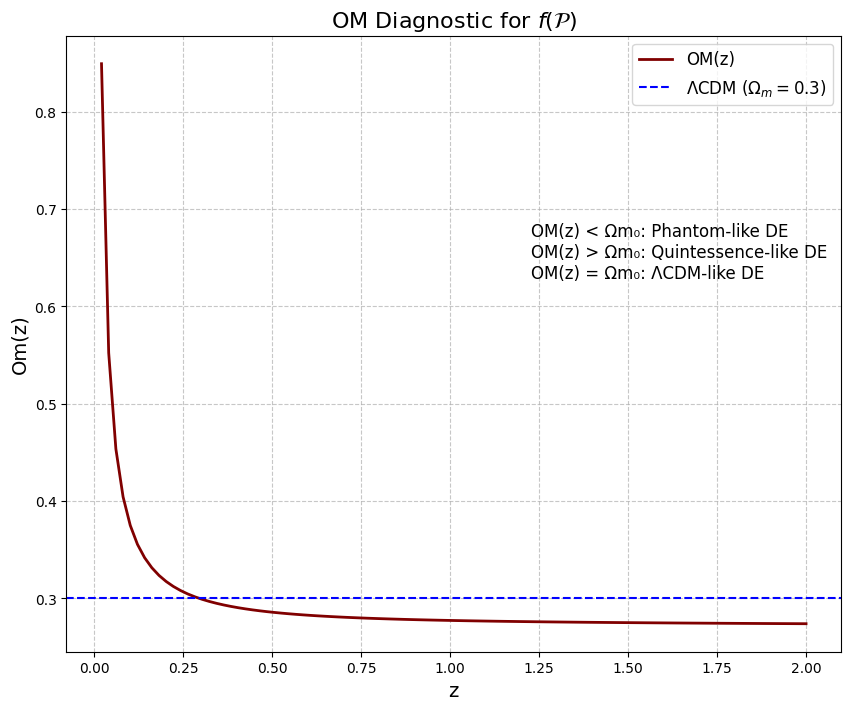}
        \caption{}
        \label{fig:om_f(P)}
    \end{subfigure}
    \begin{subfigure}[b]{0.45\textwidth}
        \includegraphics[width=\textwidth]{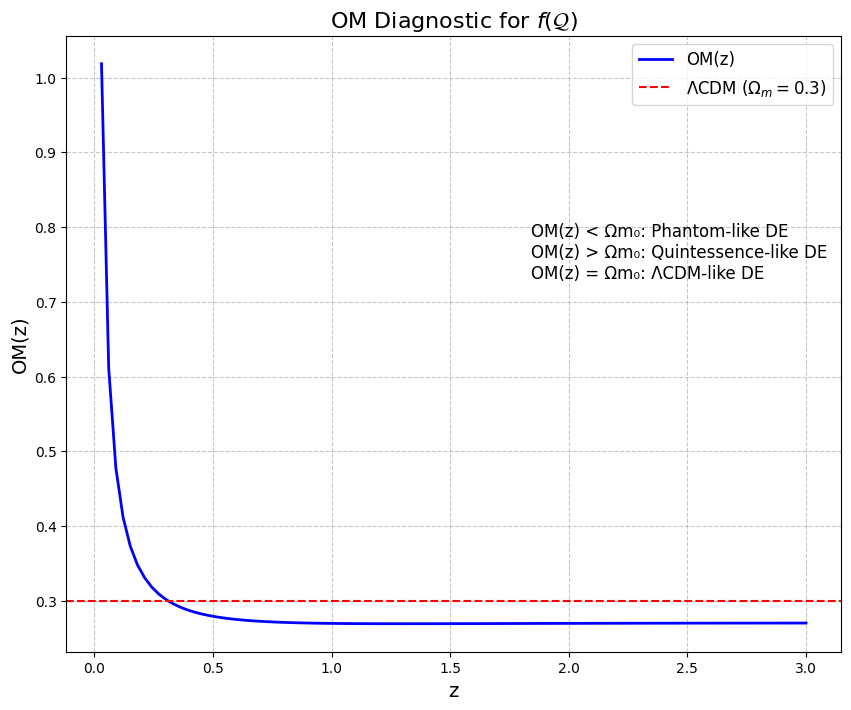}
        \caption{}
        \label{fig:om_f(Q)}
    \end{subfigure}
    \caption{Plot of evolution trajectory of $Om(z)$ diagnostic parameter for the fig: (a) $f(\mathcal{P})$ gravity and fig: (b) $f(\mathcal{Q})$ gravity.}
    \label{fig:om_f(P)_f(Q)}
\end{figure}

{\bf Om(z) diagnostic:} The figure.~\ref{fig:om_f(P)_f(Q)} presents the Om diagnostic for our cosmological models $f(\mathcal{P})$ and $f(\mathcal{Q})$, plotted against redshift z. This diagnostic tool provides in-depth revelations regarding the characteristics of dark energy in our model compared to the standard ΛCDM paradigm. Both models demonstrate dynamic behaviour across cosmic history. At high redshifts [$(z > 0.25)$ for $f(\mathcal{P})$ and $(z>0.5)$ for $f(\mathcal{Q})$], the $Om(z)$ function asymptotically approaches a value slightly below the ΛCDM reference line. This suggests that in the early universe, our model exhibits behaviour close to, but distinctly different from, the standard ΛCDM model. As we move towards lower redshifts $(z < 0.5)$, we observe a significant deviation from ΛCDM. The $Om(z)$ function rises sharply, crossing the ΛCDM reference line and continuing to increase. This behaviour indicates a transition in the nature of dark energy in our model:
\begin{itemize}
    \item At low redshifts, where $Om(z) > \Omega_m$, our models suggest quintessence-like dark energy behaviour.
    \item At very high redshifts, $Om(z) < \Omega_m$ and the models suggest the phantom-like dark energy. In the intermediate redshift as $Om(z)$ increases rapidly, the models transition into a quintessence-like dark energy regime.
\end{itemize}
It suggests a dynamic dark energy component that becomes increasingly dominant at late times, potentially offering a clear vision of the accelerated expansion of our universe and the nature of dark energy.

{\bf Other Cosmological parameters:} From the table.~\ref{tab:MCMCresults} it can be noted that the value of the matter density parameter at the current epoch is found to be $\Omega_{m0} = 0.2815^{+0.01}_{-0.02}$ for the $f(\mathcal{P})$ gravity model and $\Omega_{m0}=0.2752^{+0.02}_{-0.02}$. Both the values are less than the current Planck observation \cite{aghanim2020planck}. The value of the equation of state parameter for matter, denoted as $\omega_m$, is very close to zero.. The negative sign of $\omega_m$ indicates the presence of some exotic type of matter if the background geometry is taken to be $f(\mathcal{P})$ or $f(\mathcal{Q)}$.  The constrained value of the Barrow exponent $\Delta$ is seen to have value $0.4280^{+0.40}_{-0.27}$ for $f(\mathcal{P})$ gravity model and $0.4464^{+0.39}_{-0.30}$ for $f(\mathcal{Q})$ gravity model. Both the values are well deviated and large from the $\Delta$ value obtained in a flat universe with Einstein's gravity taken as the background geometry \cite{anagnostopoulos2020observational}. Interestingly, it seems that the $\Delta$ values for our model have a closer agreement with that of a non-flat universe with negative curvature value \cite{adhikary2021barrow}. From the above data in table.~\ref{tab:MCMCresults}, the value of the constant parameter $C$ is estimated to be roughly in the order of $\approx 10^4$ for both models. The age of the universe is calculated using the MCMC code and obtained as $t_{age} = 13.95^{+0.16}_{-0.14}$ Gyr for $f(\mathcal{P})$ gravity model and $t_{age}=13.97^{+0.14}_{-0.14}$ Gyr for $f(\mathcal{Q})$ gravity model. Both values are close enough to the current calculated age of the universe from the observational data\cite{refId0}.

\section{Discussions}\label{Sect:Conclusion}
In summary, we have initially developed the (m,n)-type model of Barrow holographic dark energy. Then we provide a precise overview of the fundamental formulations in extended cubic gravity $f(\mathcal{P})$ and symmetric $f(\mathcal{Q})$ teleparallel gravity. After that, we have equated the energy densities of the $f(\mathcal{P})$ and $f(\mathcal{Q})$ gravities to the density of the $(m,n)$-type Barrow holographic dark energy to reconstruct these functions. The reconstructed functions are found to be analytic everywhere except at some finite number of points. Thus, it is feasible to reconstruct the $f(\mathcal{P})$ and $f(\mathcal{Q})$ gravities corresponding to the $(m,n)$-type Barrow holographic dark energy framework. Next, we plotted the functions $f(\mathcal{P})$ and $f(\mathcal{Q})$ with $\mathcal{P}$ and $\mathcal{Q}$ and showed that $f(\mathcal{P})$ increases with $\mathcal{P}$ whereas $f(\mathcal{Q})$ is a decreasing function of $\mathcal{Q}$. Then, we run the MCMC to constrain the model parameters using 57 data points of combined CC and BAO datasets. The posterior distributions at $1\sigma$ and $2\sigma$ confidence levels are presented in fig.~\ref{fig:getdist_f(P)_f(Q)}. The evolution of Hubble parameter shows the model deviates from $\Lambda$CDM model at higher redshifts, though at lower redshifts they are in complete agreement. Next, we have analysed the growth of our universe through various cosmological parameters, namely deceleration parameter $q$, Equation of State parameter $\omega_{DE}$, square of the velocity of sound $v_s^2$, density parameter $\Omega(z)$ and cosmographic parameters. The graph of the deceleration parameter shows that the universe is presently undergoing accelerated expansion. The transition from a decelerating to an accelerating phase happens around $z_t \approx 0.7$ for both newly reconstructed $f(\mathcal{P})$ and $f(\mathcal{Q})$ gravity models. The EoS parameter $\omega_{DE}$ vs. redshift $z$ plot shows the existence of quintessence like dark energy for both $f(\mathcal{P})$ and $f(\mathcal{Q})$ gravity when it is constructed from $(m,n)$-type Barrow holographic dark energy. The square of the velocity of sound $v_s^2$ vs. redshift $z$ depicts the reconstructed $f(\mathcal{P})$ cubic gravity is partially stable under perturbation whereas the $f(\mathcal{Q})$ teleparallel gravity is classically stable against perturbation for the entire range of redshift. The jerk parameter $j$ is positive for both $f(\mathcal{P})$ and $f(\mathcal{Q})$, indicating an accelerated expansion phase. The density parameter shows the transition from matter dominated universe to the dark energy dominated universe as the universe evolves.\\
The $\omega'_{DE}-\omega_{DE}$ plane shows freezing region for both reconstructed $f(\mathcal{P})$ and $f(\mathcal{Q})$ gravity indicating growth of accelerated expansion of the universe. The statefinder parameter $(r,s^*)$ reveals that the dark energy was of quintessence type in the past epochs, then it passed through the $\Lambda$CDM point, and then it is predicted to evolve towards a Chaplygin gas (phantom) type in the future for both reconstructed $f(\mathcal{P})$ and $f(\mathcal{Q})$ gravity models. The $(r,q)$ parameter plot shows the existence of accelerated expansion at later epochs of the universe. From the $Om(z)$ diagnostic plot, it is apparent that the dark energy becomes quintessence type at lower redshifts. In the field of cosmology, the concept of quintessence infers a scenario where dark energy exhibits dynamic behaviour shown by a scalar field Lagrangian, and the Chaplygin gas region represents an alternative regime where dark energy is described as a perfect fluid, characterised by an equation of state similar to that of Chaplygin gas. We deploy the Akaike Information Criterion (AIC), Bayesian Information Criterion (BIC), and Deviance Information Criterion (DIC). The AIC is defined by \[
\text{AIC} = -2 \ln(L_{\text{max}}) + 2k + \frac{2k(k+1)}{N_{\text{tot}} - k - 1}\]where \( L_{\text{max}} \) is the maximum likelihood, \( k \) is the number of parameters, and \( N_{\text{tot}} \) is the total number of observations.The BIC is given by, \[\text{BIC} = -2 \ln(L_{\text{max}}) + k \ln(N_{\text{tot}})\] where \( L_{\text{max}} \) is the maximum likelihood, \( k \) is the number of parameters, and \( N_{\text{tot}} \) is the total number of observations. The Deviance Information Criterion (DIC) is given by:
\[\text{DIC} = \bar{D} + p_D\] where:\[\bar{D} = \frac{1}{S} \sum_{i=1}^S D(\theta_i)\]is the mean deviance across the posterior samples, and \( p_D \) is the effective number of parameters, calculated as: \[p_D = \bar{D} - D(\hat{\theta})\]where \( D(\hat{\theta}) \) is the deviance at the posterior mean of the parameters. We have performed a comparison of AIC, BIC, and DIC between our models and the $\Lambda$CDM model. The comparison data is provided in table.~\ref{tab:model_metrics_comparison}. The $\Delta$DIC is less than 2 for both the models. From the $\Delta$AIC data, it is evident that both the models are in moderate tension with $f(\mathcal{Q})$ having the lower value. The $\Delta$BIC data shows that the $f(\mathcal{Q})$ model has a lower value compared to another model. Although the ΛCDM model shows the best fit with the lowest AIC, our combined AIC, BIC, and DIC results suggest that all the tested models are viable. This implies that none of the models can be rejected based solely on the current data. The reduced chi-square value of 0.62317065 and 0.61703767 suggests that the model provides a good fit to the data, with a value less than 1 indicating that the model is not over fitting the data. Future research in the reconstructions of \( f(\mathcal{P}) \) and \( f(\mathcal{Q}) \) gravity from $(m,n)$-type Barrow holographic dark energy can explore several promising avenues. One intriguing direction is the introduction of a 4-parameter generalised entropy, such as the Nojiri-Odintsov-Paul entropy \cite{Nojiri_2022}, which reduces to Barrow entropy under specific parameter limits, thereby reproducing the proposal of this work with a broader theoretical framework. Another key aspect worth investigating is the possibility that the Barrow exponent \( \Delta \) might vary with energy scale \cite{Di_Gennaro_2022}, offering novel implications for early universe cosmology and the potential consideration of a time-varying \( \Delta \), though this diverges from the holographic dark energy context addressed here. Additionally, incorporating more extensive cosmological datasets, including CMB, Pantheon+, and Planck data, could provide tighter constraints on model parameters, even though this extends beyond the scope of the present work. Lastly, running simulations of large-scale structure (LSS) formation using this model would significantly enhance our understanding of the universe's evolution, particularly in contrast to the standard \(\Lambda\)CDM paradigm.\\

{\bf ACKNOWLEDGEMENTS:} 
AK is thankful to IIEST, Shibpur, India for Institute Fellowship
(JRF).\\

{\bf Data Availability Statement:}
All data generated or analysed during this study are included in this published article. No new data were created or analysed in this study.\\

{\bf Conflict of Interest Statement:}
The authors declare that the research was conducted in the absence of any commercial or financial relationships that could be construed as a potential conflict of interest.

\bibliographystyle{apsrev4-2}

\bibliography{references}

\end{document}